\documentclass[12pt,preprint]{aastex}
\usepackage[onecolumn,numberedappendix]{emulateapj5}

\usepackage{epsfig}
\usepackage{rotating}
\tightenlines
\slugcomment{Accepted to the Astrophysical Journal}

\newcommand     \cm     {\,{\rm cm}}

\newcommand     \kms    {\,{\rm km~s}^{-1}}

\newcommand     \yr     {\,{\rm yr}}

\newcommand{\smyr}{{ M_\odot\ \rm yr^{-1}}}
\newcommand{\sm}{{ M_\odot}}
\newcommand{\cth}{c_{\rm th}}

\newcommand{\beq}{\begin{equation}}
\newcommand{\eeq}{\end{equation}}
\newcommand{\beqa}{\begin{eqnarray}}
\newcommand{\eeqa}{\end{eqnarray}}

\newcommand{\eee}{$^{-3}$}

\newcommand{\krho}{{k_\rho}}

\newcommand{\mds}{\dot m_*}
\newcommand{\mdsd}{\dot m_{*d}}
\newcommand{\msd}{m_{*d}}

\newcommand{\mcore}{M_{\rm core}}

\newcommand{\esd}{\epsilon_{*d}}

\newcommand{\htwo}      {H$_2$}
\newcommand{\nh}        {n_{\rm H}}

\newcommand{\urad}      {{u}_{\rm rad}}

\newcommand{\vecF}      {{\bf F}}

\newcommand{\vecnabla}  {{\bf\nabla}}

\newcommand{\mh}        {m_{\rm H}}
\newcommand{\pbpt}      {\frac{\partial}{\partial t}}
\newcommand{\epsth}     {\epsilon_{\rm th}}
\newcommand{\epsnucd}   {\dot \epsilon_{\rm nuc}}
\newcommand{\enucd}     {\dot E_{\rm nuc}}

\newcommand{\fint}      {F_{\rm int}}

\newcommand{\rsp}       {r_{\rm sp}}
\newcommand{\spr}       {{\bf \sigma'}}
\newcommand{\vff}       {v_{\rm ff}}
\newcommand{\vffs}      {v_{\rm ff,*}}
\newcommand{\vkep}      {v_{\rm Kep}}
\newcommand{\vrot}      {v_{\rm rot}}
\newcommand{\vecg}      {{\bf g}}
\newcommand{\vecv}      {{\bf v}}

\newlength{\figwidth}
\countdef\decade=200
\advance\decade by \year
\countdef\hours=201
\advance\hours by \time
\divide\hours by 60
\countdef\mins=202
\advance\mins by \hours
\multiply\mins by 60
\multiply\hours by 100
\countdef\miltime=203
\advance\miltime by \hours
\advance\miltime by \time
\advance\miltime by -\mins

\begin{document}

\title{The Formation of the First Stars I.\\ 
Mass Infall Rates, Accretion Disk Structure 
and Protostellar Evolution}
        %\\
        %{\small DRAFT: \today}}

\author{Jonathan C. Tan$^1$ and Christopher F. McKee$^2$
        }
\affil{1. Princeton University Observatory, Peyton Hall, Princeton, NJ
08544, USA.\\jt@astro.princeton.edu}
\affil{2. Departments of Physics \& Astronomy, University of California,
Berkeley, CA 94720, USA.\\cmckee@mckee.berkeley.edu}

\begin{abstract}
  We present a theoretical model for primordial star formation. First
  we describe the structure of the initial gas cores as virialized,
  quasi-hydrostatic objects in accord with recent high resolution
  numerical studies. The accretion rate can then be related to
  characteristic densities and temperatures that are set by the
  cooling properties of molecular hydrogen. We allow for rotation of
  the gas core, assuming angular momentum conservation inside the
  sonic point of the flow. In the typical case, most mass then reaches
  the star via an accretion disk.  The structure of the inner region
  of this disk is described with the standard theory of viscous disks,
  but with allowance for the substantial energies absorbed in ionizing
  and dissociating the gas. The size of the protostar and its
  luminosity depend upon the accretion rate, the energetics of the
  accreting gas, and the ability of the radiation to escape from the
  stellar accretion shock. We combine these models for the infall
  rate, inner disk structure, and protostellar evolution to predict
  the radiation field that is the basis for radiative feedback
  processes acting against infall (Paper~II). 
  For realistic initial angular momenta, the photosphere of the
  protostar is much smaller and hotter than in the spherical case,
  leading to stronger radiative feedback at earlier stages in the
  evolution. In particular, once the star is older than its
  Kelvin-Helmholtz time, contraction towards the main sequence causes
  a rapid increase in ionizing and far-ultraviolet luminosity at
  masses $\sim 30\sm$ in the fiducial case.  
  Since the cores out of which the first stars  formed were much more massive
  than $30\sm$ and since feedback is dynamically unimportant at 
  lower masses, we
  conclude that the first stars should have had masses $\ga 30\sm$.
\end{abstract}


\section{Introduction\label{S:intro}}

The formation of the first stars marks the beginning of the long saga
of galaxy formation and evolution. These stars, if relatively massive,
are thought to bring the Universe out of the ``dark ages'', heating
and reionizing the intergalactic medium, and enriching their
surroundings with metals. The stellar mass determines whether or not
supernovae occur, and if so the energy of the explosion, composition
of the ejecta, and type of remnant. Very energetic supernovae may
produce gamma-ray bursts (GRBs). The first stars may have had a
profound influence on the formation of supermassive black holes, 
globular clusters, and proto-galactic fragments.
For all these reasons, as well as simple curiosity about
the very first stars in the Universe, we wish to understand the
process of primordial star formation and how, in particular, the
formation process may control the resulting stellar mass.

When and how quickly the universe was reionized depends on the
clumpiness of the intergalactic medium (IGM), the time of first star
formation, its efficiency, and the ionizing luminosity per stellar baryon
(e.g. Madau, Haardt, \& Rees 1999). The latter depends on the initial
mass function (IMF) of the stellar population (Tumlinson \& Shull 2000;
Bromm, Kudritzki, \& Loeb 2001; Ciardi et al. 2001;
Schaerer 2002), although Bromm et al. (2001) showed that once the
stars all have masses $\ga 300\sm$, the spectral luminosity per unit
stellar mass becomes almost independent of the IMF. The hardness of the
radiation field is also somewhat sensitive to the IMF
(e.g. Tumlinson \& Shull 2000; Schaerer 2002), so that He reionization
can be affected. Significant reionization may also result from
a population of supermassive black holes, making it more
difficult to draw conclusions on the nature of the stellar population.
%For example, a Salpeter IMF of zero metallicity stars with $d{\cal
%N}_*/d m_*\propto m_*^{-2.35}$, ranging from 0.1 to 100~$\sm$,
%produces about 4000 Lyman-continuum (Ly-c) photons per stellar baryon
%over its lifetime. An IMF consisting of only very massive stars
%($m_*\ga 100\sm$, produces $\sim 10^5$ Ly-c photons per baryon. If
%it takes $\sim 10$ Ly-c photons per baryon to reionize the Universe
%(ref??), then for the above two cases, the minimum cosmic density of
%early star formation for reionization is $0.25\%$ and $0.01\%$,
%respectively.
The recent microwave background polarization results reported by Kogut
et al. (2003) imply a relatively early epoch of reionization and a
very top-heavy IMF for models invoking reionization due to stars.

As with contemporary stars, the bolometric luminosities and the total
energy output of the first stars depend on their mass. Thus, along
with the overall efficiency of star formation, the IMF has
implications for the contribution the primordial stars make to the
near infra-red background (Santos, Bromm, \& Kamionkowski 2002; Salvaterra \& Ferrara 2003). 
Unfortunately the observational determination of this
background is hampered by the zodiacal foreground
emission.  Clusters of primordial stars may potentially be detected by
the next generation of space-based near infra-red observatories,
perhaps showing up in the number count versus flux distribution of
very faint sources, or in the fluctuations of the background
(Magliocchetti, Salvaterra, \& Ferrara 2003).

The metal enrichment of the IGM from core-collapse supernovae depends
on the mass of the stellar progenitors. Massive primordial stars are
thought to have much smaller mass-loss rates than their present-day
cousins (Barrafe, Heger, \& Woosley 2001) so the mass at core collapse
may be quite similar to the initial mass. For supernova progenitor
masses $\ga 260\sm$, collapse is thought to proceed directly to a
black hole, as the temperatures resulting from the pair-instability
collapse are great enough to photodisintegrate nuclei, which counters
explosive oxygen and silicon burning (Fryer, Woosley, \& Heger 2001;
Heger \& Woosley 2002).  There is expected to be relatively little
mass ejection and metal enrichment from such supernovae. For smaller
masses, down to about $\sim 140\sm$, explosive O and Si burning is
able to completely disrupt the star, leaving no remnant and ejecting
large quantities of heavy elements. Progenitors from $\sim 40$ to
$140\sm$ form black holes, with relatively inefficient metal ejection,
while less massive supernovae form neutron stars, with more normal
enrichment rates (however, see Umeda \& Nomoto 2003). 
In principle, metallicity determinations from high
redshift absorption line systems or from very metal poor local stars
(e.g. Aoki et al. 2002; Christlieb et al. 2002) can constrain the IMF
of the early generations of stars.

A phenomenon that may be related to supernovae and hypernovae is the
production of relativistic ejecta and thereafter a gamma-ray burst
(Woosley 1993; Paczynski 1998; MacFadyen \& Woosley 1999; Tan, Matzner
\& McKee 2001; Matzner 2003). The amount of relativistic ejecta
accelerated by a blast wave in the stellar atmosphere is a sensitive
function of the progenitor mass, density structure, and explosion
energy (Tan et al. 2001). Massive stars that mostly collapse to a
black hole and impart a large energy to the small fraction of ejected
mass, are excellent accelerators of relativistic ejecta.  However, for
efficient gamma-ray production this ejecta should either interact with
a relatively dense circumstellar envelope, or be subject to internal
shocks. If GRBs and associated afterglows are found at high redshift
they will provide excellent probes of the IGM, for example through the
study of 21~cm absorption lines (Furlanetto \& Loeb 2002).

The impact of the first stars on the heating and metal enrichment (and
thus cooling) of the IGM may have important implications for the
formation of other objects, such as supermassive black holes, globular
clusters, and galaxies.  If the first stars were sufficiently massive
to collapse efficiently into black holes and their feedback on the
surrounding gas was relatively weak, then they might become the seeds
that grow to millions of solar masses and more.  Other formation
scenarios are possible (e.g. Rees 1984).  One alternative requires
direct collapse of gas to a black hole without a nuclear-burning
stellar phase (e.g. Baumgarte \& Shapiro 1999), and applies to
``stars'' with $m_*\ga 10^6\sm$ (Zeldovich \& Novikov 1971). However,
concentrating such a large mass of gas into a hydrostatic structure,
without first forming less massive protostars, would be difficult for
the very first generation of stars since molecular cooling acts to
reduce the Jeans mass (e.g. Abel, Bryan, \& Norman 2002, hereafter
ABN). This scenario could perhaps be resuscitated for subsequent
generations of star formation if the FUV background radiation produced
by earlier generations of star formation were large enough to prevent
$\rm H_2$ formation (Bromm \& Loeb 2003).  Other formation mechanisms 
have their origins in
a very dense star cluster, composed either of nuclear-burning or
compact stars. The ability of such clusters to form in the early
universe is likely to be influenced by the nature of primordial stars;
for example, if primordial stars are relatively massive, then the
resulting feedback may inhibit further star formation in the 
vicinity (Omukai \& Nishi 1999).
The same considerations apply to early globular cluster formation.
%jct note to self:check metallicity floor for globulars?
The dominant feedback on cosmological scales that affects galaxy
formation is likely to be via the intensity of the FUV background.  If
this is very high, then by suppressing the $\rm H_2$ abundance, the
collapse of baryons into ``mini-halos'' 
%cfm Dwarf galaxies are massive enough that they are unaffected
%    by $T~10^4 K$: and the formation of dwarf galaxies 
%cf
would be inhibited.

A description of primordial star formation requires relatively simple
physics and chemistry, mainly because, by definition there are no
complicated feedback processes from earlier stellar
generations. For most early generations of stars, the only feedback
variable will be the intensity of the FUV background, with perhaps
some additional influence from hard X-rays (Cen 2003).  In contrast to the
present-day case, dynamically-important magnetic fields and dust
grains are probably not present during the initial stages of collapse.
The initial conditions for star formation are well-defined
cosmological fluctuations. Given this relative simplicity, we have
greater confidence in the applicability of the results of numerical
simulations that have followed the collapse of cosmological scale
perturbations down to almost stellar dimensions (ABN; Bromm, Coppi, \&
Larson 1999; 2002, hereafter BCL). This confidence is strengthened by the
fact that it appears to be the microphysics of $\rm H_2$ cooling that
determines the types of baryonic structures that are formed, and not,
for example, the details of the initial power spectrum of fluctuations
in dark matter density. Molecule formation was predicted to occur in
the presence of residual electrons left over from recombination via
$\rm H^-$, (McDowell 1961; Hirasawa, Aizu, \& Taketani 1969).
Including this physics, the results of the recent numerical
simulations suggest that the initial gas cores out of which stars form
are quite massive, $\mcore \sim 100 - 1000\sm$, in agreement with
earlier theoretical expectations (Yoneyama 1972; Hutchins 1976; Silk
1977; Carlberg 1981).  However, at very high densities three-body
molecule formation becomes efficient, and it was speculated that the
increased cooling would lead to much smaller Jeans masses $\sim
M_\odot$ (Palla, Salpeter, \& Stahler 1983). The simulations indicate
that this process does occur, but produces only a single initially sub-solar mass
protostar at the center of the much larger core. Accretion then builds
up the mass of the star (Omukai \& Nishi 1998, Ripamonti et al. 2002).

The goal of this paper is to model the growth of the protostar from
very small masses to large, and to determine the total energy output
from the star. Here we do not consider how feedback processes may
inhibit the accretion to the protostar, deferring this important
question to other papers. In \S\ref{S:structureform} we briefly
review the stages of the collapse immediately prior to protostellar
core formation. We then describe the accretion rate to and the
structure of the infall envelope around the protostellar core and disk
(\S\S \ref{S:accretion}, \ref{S:infall}).   We model the
inner accretion disk in \S\ref{S:disk} and the evolution of the
protostar in \S\ref{S:evol}.  The results for the bolometric and
ionizing luminosities are presented in \S\ref{S:lum}. 
%cfm
Appendix A describes the accretion of energy by a protostar and
its disk. General results on the
structure and luminosity
of an accreting protostar are given in Appendix B.
%cf
Radiative
feedback processes are considered in Paper II, while magnetic field
generation and hydromagnetic winds have been modeled by Tan \&
Blackman (2004).

\section{Overview of Baryonic Structure Formation: from Cosmological
Fluctuations to Star-Forming Cores}\label{S:structureform}

A simple analytic picture of baryonic structure formation is the
following (e.g. Peebles 1993; Tegmark et al. 1997; Madau 2002).
Recombination at a redshift $z\simeq 1200$ heralds the start of the
``dark ages''.  Thermal equilibration of matter and radiation is
maintained via the residual free electrons until $z\sim 160$. Until
this point the cosmological Jeans mass (gas plus dark matter), $M_J
\propto (T/\rho^{1/3})^{3/2}$, is therefore independent of $z$ and has
typical value $\sim 10^5\sm$, similar to the scale of globular
clusters. At lower redshifts, baryons cool adiabatically so that
$T\propto (1+z)^2$ and $M_J\propto (1+z)^{3/2}$. Gas collects in dark
matter halos and cools further via $\rm H_2$ line cooling. At some
point the first luminous objects form, reionize the Universe to a
temperature of $\sim 10^4\:{\rm K}$ and raise the Jeans mass to
galactic scales of $\sim 10^{9-10} [(1+z_{\rm re-ion})/10]^{-3/2}\sm$.

Numerical simulations (ABN, BCL) have largely confirmed this picture.
Collapse is seen to proceed along filaments, with objects of
characteristic mass $M_J$ forming and rapidly merging to build-up
somewhat larger masses. The simulations of ABN use adaptive
mesh-refinement and follow cosmological perturbations in a relatively
small volume from $z=100$ to $z\simeq 20$, where the first luminous
object forms. These simulations are stopped when molecular lines, the
main coolant at this stage, become optically thick. Spherically
symmetric simulations (Omukai \& Nishi 1998; Ripamonti et al. 2002)
are able to follow the collapse all the way to the formation of a
hydrostatic protostellar core.

Summarizing the above numerical results: a pre-galactic halo (gas plus
dark matter) of $\sim 10^{5-6}\sm$ forms at the intersection of
several filaments; a quasi-hydrostatic, gas-dominated core forms with
mass $\sim 4000\sm$, $r\sim 10\:{\rm pc}$, $T\sim 200-300\:{\rm K}$ (set
by molecular cooling, with the mass fraction of molecules $\sim
10^{-3}$ and their formation catalysed by free electrons via $\rm
H^-$); gradual contraction of the core is driven by cooling in the
dense interior; rapid 3-body $\rm H_2$ formation occurs at $n\ga
10^{8}\:{\rm cm^{-3}}$ so that most H is molecular by the time
densities reach $\sim 10^{11}\:{\rm cm^{-3}}$, leading to strong
cooling and supersonic inflow; and a hydrostatic core of mass $\simeq
5\times 10^{-3}\sm$ and radius $\sim 14\:{\rm R_\odot}$ forms when the
gas becomes optically thick to the cooling radiation (continuum
cooling from collision-induced absorption).

\section{The Accretion Rate: Isentropic Accretion Model}
\label{S:accretion}

Consider a  gas cloud that is initially 
in approximate hydrostatic equilibrium. We assume that
this cloud undergoes gravitational collapse, 
forming one or more protostars.  We focus on the gas
that will go into a single protostar (or possibly a binary);
we term this the protostellar core.  We assume that the
core is approximately spherical,
with a baryonic density $\rho(r)$ and mass $M(r)$.
Since the core is initially in approximate hydrostatic
equilibrium, it will undergo an ``inside-out" collapse
(Shu 1977), in which the protostellar mass $m_*$ grows with time
from a very small initial value to its final value.

     In their study of contemporary star formation,
McKee \& Tan (2002, 2003) assumed that
the mass of any protostellar accretion disk was small compared
to the stellar mass.
The high accretion rates and possible absence
of magnetic fields inferred for primordial
star formation imply that, once most accretion is via a 
disk, then the rate limiting step for accretion is likely to be 
gravitational torques in the disk. 
Such torques become efficient only after the disk
becomes relatively massive (disk mass $m_d\simeq \frac 13 m_*$)
and self-gravitating
(Adams, Ruden, \& Shu 1989; Shu et al. 1990; Gammie 2001).
Let 
\beq
m_{*d}=m_*+m_d\equiv(1+f_d)m_*
\eeq
be the mass of the protostar and its associated accretion disk,
where $f_d$ is
the ratio of the disk to stellar mass.
We adopt $f_d\simeq 1/3$ as a fiducial value. 
We allow for the possibility that 
only a fraction $\esd$ of the mass in the core
winds up in the protostar plus disk due to 
protostellar outflows.  
At any time after the collapse
has commenced, we can then associate a core mass $M(r)=\msd/\esd$,
and hence a density $\rho(r)$, with the protostar and disk when
its mass is $\msd$.
In contemporary star formation, the main driver for protostellar outflows
is believed to be hydromagnetic winds (K\"onigl \& Pudritz 2000; Shu
et al. 2000). In this paper we assume that 
magnetic fields, and therefore protostellar outflows,
are negligible in primordial star formation, so that $\esd=1$;
however, we shall keep $\esd$ in the
equations for generality. (Tan \& Blackman 2004 consider the possibility
of magnetic field generation by a disk dynamo, and the 
influence of the resulting
outflow on $\esd$.)

In general, 
the accretion rate to the protostar and any associated
accretion disk can be expressed as 
\beq 
\mdsd=\phi_*\;\frac{\msd}{t_{\rm ff}},
\label{eq:mds1}
\eeq 
extending the expression given by 
Stahler, Shu, \& Taam (1980) and McKee \& Tan (2002, 2003)
to the case in which the disk mass is a significant fraction
of the protostellar mass.  Here
$\phi_*$ is a numerical parameter of order unity and $t_{\rm ff}
=(3\pi/32G\rho)^{1/2}$ is the free-fall time measured at the
location 
in the initial core where the interior baryonic mass is
$M=m_{*d}/\esd$. We assume that the mass fraction 
of dark matter is small, which is valid 
$\la 0.3\:{\rm pc}$ 
from the core center 
(ABN; note, however, that they assumed a baryon to
dark matter ratio of 0.064, compared to 0.21 derived from the results
reported by Spergel et al. 2003, so their results overestimate
the effect of dark matter). 
On larger scales, the effect of dark
matter is to reduce the free-fall time, or equivalently increase the
value of $\phi_*$, by factors of order unity relative to the purely
baryonic case.  

Following McKee \& Tan (2002), we now assume that
the core out of
which the star is forming is in a self-similar virial equilibrium with
a density $\rho(r)\propto r^{-\krho}$
and a pressure $P\propto r^{-k_P}$. The pressure then
obeys the polytropic relation
\beq
P=K_p\rho^{\gamma_p},
\eeq
with $K_p={\rm const}$ and $\gamma_p=k_P/\krho$.  In hydrostatic equilibrium,
$k_P=2(\krho-1)$, so that $\gamma_p=2(1-1/\krho)$.  
If the gas changes in time, 
e.g., if it undergoes a reversible compression or rarefaction,
then the change in pressure
is related to that in the density by the adiabatic
index $\gamma=\delta\ln P/\delta\ln \rho$.
If we define the parameter $K$ by $P\equiv K\rho^\gamma$, then
$K$ is a function of the entropy 
and is therefore termed the ``entropy parameter" (McKee
\& Holliman 1999).
If the adiabatic and polytropic
indexes are equal ($\gamma=\gamma_p$), then $K=K_p$:
a polytropic gas is isentropic and vice versa; furthermore,
if the gas is initially polytropic, 
it will remain so under adiabatic compression or expansion. 

   McLaughlin \& Pudritz (1997) determined the accretion rate for
a collapsing isentropic, singular polytropic sphere, and based on
their results McKee \& Tan (2002) showed that in this case
$\phi_*\simeq 1.62-0.48\krho$. If the gas is not isentropic
($\gamma\neq\gamma_p$), the value of $\phi_*$ should not change
significantly provided that the
gas is convectively stable ($\gamma_p<\gamma$) and
is not stiff ($\gamma < \frac 43$).
For the case of primordial star formation the
assumption that the gas is isentropic should be quite good:
BCL and ABN show that
H$_2$ cooling causes the gas to accumulate
at a density of order $10^4$ cm\eee\ and a temperature of
a few hundred K, which sets the value of the entropy; Omukai
\& Nishi (1998) have shown that the subsequent compression of
the gas obeys $\gamma\simeq 1.09$, leading to an isentropic
gas distribution with $\gamma\simeq \gamma_p\simeq 1.1$.

     McKee \& Tan (2002, 2003) expressed the protostellar
accretion rate in terms of the mass of the star and
the pressure at the surface of the core out of which the
star is forming.  However, the accretion rate can equally
well be expressed in terms of the entropy parameter $K$ (Yahil 1983).
From the equation of hydrostatic equilibrium, one
can readily show that
for an isentropic gas with $\gamma=\gamma_p$,
\beq
\rho=\left[\frac{(3-\krho)k_P^3 K^3}{4\pi G^3M^2}\right]
        ^{1/(4-3\gamma_p)},
\label{eq:rho}
\eeq where $M$ is the mass of the core interior to the point at which
the density is $\rho$.  
Using equation (\ref{eq:rho}) in equation
(\ref{eq:mds1}), we then find 
\beq 
\mdsd=\frac{8\phi_*\esd}{\surd
        3}\left[\frac{(3-\krho)k_P^3 K^3}{2(2\pi)
        ^{5-3\gamma_p}G^{3\gamma_p-1}}\right]^{1/[2(4-3\gamma_p)]}
        M^j,
\label{eq:mds2}
\eeq
where
\beq
j\equiv 3\left(\frac{1-\gamma_p}{4-3\gamma_p}\right).
\eeq
For $m_{*d}\propto M$ (actually, we are assuming $m_{*d}=M$),
this corresponds to a time dependence $\mdsd\propto t^{j/(1-j)}$.
Thus, whereas contemporary star formation accelerates
($\gamma_p<1$ so that $j/[1-j]>0$---McLaughlin \& Pudritz 1997;
McKee \& Tan 2002, 2003), primordial star formation {\it decelerates}
($4/3>\gamma_p>1$ so that $j/[1-j]<0$) (Omukai \& Nishi 1998).

        For $\gamma\simeq 1.09$ (Omukai \& Nishi 1998),
we have $k_\rho\simeq 2.2$, so that $\mdsd\propto
M^{-0.37}$. The dependence of $\mdsd$ on $M$ is quite sensitive to
$\gamma_p$. Ripamonti et al. (2002) followed the growth of the
protostar up to about $0.5\sm$, and their result of $\mdsd\propto
M^{-0.522}$, i.e. $j=-0.522$, implies an effective value of
$\gamma_p=1.114$.  We shall adopt an intermediate value $\gamma_p=1.1$
for the entire collapsing
cloud. With this value we have $k_\rho = 20/9$ and $k_P=22/9$.

       The normalization of the accretion rate then depends on two
parameters, $K$ and $\phi_*$. First consider the entropy
parameter $K$.
As remarked above, ABN and BCL show that before the
gas can collapse, it passes through a stage when its density is about
$10^4$~cm\eee, which is the density at which \htwo\ is
thermalized, and a temperature of about 200~K, which is the minimum
temperature to which \htwo\ can cool the gas. However, the core can also
be supported by turbulent motions: in the simulation of ABN it is
permeated by a cascade of weak shocks, whose velocity is about the sound
speed (T. Abel, private communication, 2002). The total pressure is
$P\equiv \rho c^2 = \rho (c_{\rm th}^2 + \sigma_{\rm turb}^2)$, where $c$
is the effective sound speed, $c_{\rm th}$ is the thermal sound speed,
and $\sigma_{\rm turb}$ is the 1-D velocity dispersion of the turbulent
motions. In this case we take $\sigma=c_{\rm th}/\sqrt{3}$, so that the
total pressure is a factor 4/3 higher than the thermal pressure.
Equivalently we may define an effective temperature $T_{\rm eff}=P/(nk)$,
that is a factor 4/3 greater in this case.
We consider primordial gas with $X=0.76$ and $Y=0.24$ so that in the
atomic phase $\mu=1.22 m_{\rm H}$ and $n_{\rm He}=0.079 n_{{\rm H}}$. The
isothermal sound speed is then $\cth=\sqrt{kT/\mu}= 1.425 (T/300\:{\rm
K})^{1/2} \kms$.  
Normalizing the entropy parameter to these values gives
\begin{eqnarray}
K&=&1.88\times 10^{12}\left(\frac{T_{\rm eff}}{300~{\rm
               K}}\right)\left(\frac
           {10^4~{\rm cm}^{-3}}{\nh}\right)^{0.1}~~~{\rm cgs},\\
           &\equiv& 1.88\times 10^{12}K'~~~{\rm cgs}.
\end{eqnarray}

The parameter $\phi_*$ depends on the type of collapse solution, which
depends on the initial condition: Collapse from a singular isothermal
sphere leads to the Shu (1977) solution for inside-out collapse with
$\phi_*=0.663$, whereas analogous collapse from a singular polytropic
sphere with $\gamma_p=1.1$ gives $\phi_*\simeq 0.55$ (McKee \& Tan
2003). 
The corresponding value of the  accretion rate 
is substantially less than that which occurs in
the collapse of a uniform core (Larson 1969; Penston 1969); for the
isothermal case, it is 48 times less
(Hunter 1977). The simulations of ABN suggest that primordial star
formation proceeds in a manner intermediate between these two
extremes, but closer to the inside-out collapse: at the time of
protostar formation the envelope is contracting at about a third of
the sound speed.
Hunter (1977) has described the family of isothermal
solutions that span the 
range between the Shu solution and the Larson-Penston solution.
In particular his model (11b) has
collapse velocities of about $v_{\rm in}\simeq \cth/3$ for the gas at
the time of protostar formation. Compared to the Shu-type solution (eq.
\ref{eq:mds2}), the accretion rate is about a factor 2.6 larger.
Assuming the same increase applies to the $\gamma_p\simeq 1.1$ case,
we find (from eq. \ref{eq:mds2})
\begin{eqnarray}
\mdsd&=&0.026 \esd K'^{15/7}\left(\frac{M}{M_\odot}\right)
             ^{-3/7}~~\smyr,\\ 
         &=& 0.026 \frac{
         \esd^{10/7}K'^{15/7}}{(1+f_d)^{3/7}}
         \left(\frac{m_*}{M_\odot}\right)^{-3/7}~~\smyr.
\label{eq:mds3}
\end{eqnarray}
Note that the stellar accretion rate and the mass flow rate
through the inner accretion disk is a factor $(1+f_d)^{-1}\rightarrow
3/4$ smaller than this. 
The accretion rate declines in time as $t^{-3/10}$.
The time required to build up a given stellar mass is
\beq
t_{*}=27 (1+f_d)^{10/7} 
     \esd^{-10/7} K'^{-15/7}
       \left(\frac{m_*}{M_\odot}\right)^{10/7}~~{\rm yr}.
\label{eq:tstar}
\eeq
The lifetimes, $t_{\rm life}$, of very massive stars have been estimated
to be $\sim 2$~Myr (Schaerer 2002). The condition $t_{*}<t_{\rm life}$
implies that a maximum stellar mass of about $2000\sm$ could be
accumulated in the absence of any feedback processes. On these scales the
effect of dark matter becomes important, so that the formation timescale
for a given baryonic mass becomes shorter, raising the above mass limit
by a modest factor.

\begin{figure}[h]
\begin{center}
\epsfig{
%mdotcomp
        file=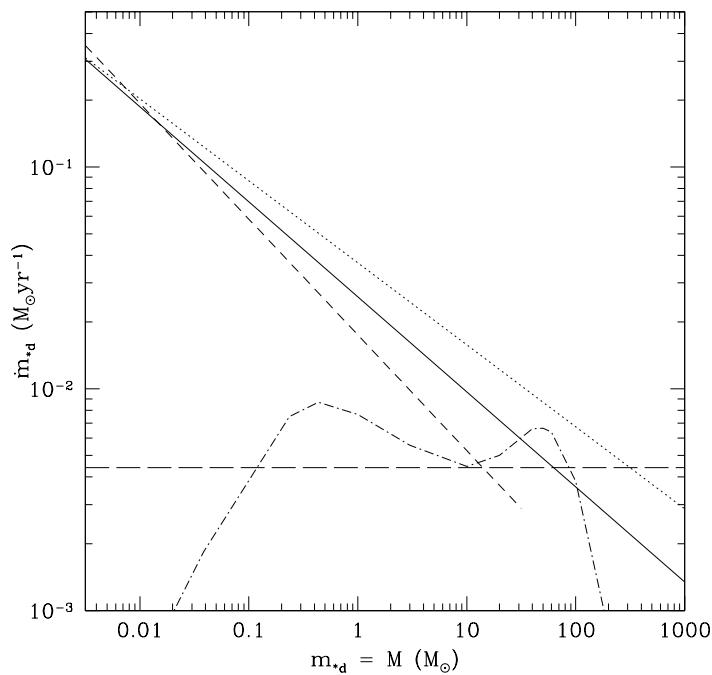,
        angle=0,
        width=\figwidth
}
\end{center}
\caption{ \label{fig:mdotcomp} Mass accretion rate onto the protostar and
  disk as a function of the current collapsed mass, $m_{*d}$ (we
  assume $\esd=1$, so that $m_{*d}=M$).  {\it Solid} line fiducial
  model (with $K^\prime=1$) from eq. (\ref{eq:mds3}). {\it Dotted}
  line from Omukai \& Nishi (1998). {\it Dashed} line is analytic
  result from Ripamonti et al. (2002). {\it Long-dashed} line is the
  constant accretion rate used in protostellar evolution models of
  Stahler et al.  (1986) and Omukai \& Palla (2001). {\it Dot-dashed}
  line is the settling inflow rate at the final stage of the
  simulation of ABN, now as a function of the enclosed mass. Note that
  the decline of this rate at small masses is due to the lack
  of the full set of high density cooling processes in the simulation
  of ABN.}
\end{figure}

The accretion rate for this model is shown in Figure
\ref{fig:mdotcomp}, along with the rates predicted by Ripamonti et al.
(2002) and Omukai \& Nishi (1998). Omukai \& Nishi assumed that the
collapse would follow the Larson-Penston solution, which is
appropriate for an initially uniform cloud.  
In the Larson-Penston solution, the infall velocity is about 
three times the sound speed at the point of protostellar core
formation.
However, the calculations of
Abel et al. (2000, 2002) show that the
protostellar core is highly centrally concentrated and has only subsonic
infall motions.  This is to be expected, since it is evolving
quasistatically due to the slow radiation losses from the trace amounts
of molecular hydrogen present in the primordial gas.  In order to fit
their observed accretion rates with an analytic model, ON were forced to
adopt an artificially small value of the entropy
parameter $K$, because of their use of the Larson-Penston
solution. Equation (\ref{eq:mds3}) is a more accurate description of how
the normalization of the accretion rate depends on the properties of the
pre-stellar core.

\subsection{The Mass Scale of Primordial Star Formation}

        The mass of the protostellar cloud can be inferred from
equation (\ref{eq:rho}):
\beq
M=543 K'^{3/2}\left(\frac{10^4\cm^{-3}}{\nh}\right)^{7/20}~~~~M_\odot,
\label{eq:M}
\eeq
where $M$ is the mass of gas at densities greater than $\nh$ under
the assumption that most of the mass is polytropic.  This mass
is somewhat less than the maximum possible mass that can be
gravitationally stable for these values of $\gamma_p$ and $\nh$,
as is generally true for polytropes (see Fig. 6 in McKee \& Holliman
1999).

        We can estimate the mass of the molecular core in this cloud
by equating the formation time for H$_2$ to the dynamical time,
which we take to be $t_*$. Rapid H$_2$ formation can occur only
at high densities via a three-body process.
The rate constant, $k_{\rm 3b}$, for $\rm H + H + H \rightarrow H_2 +
H$ used in the simulation of ABN was $k_{\rm 3b}=1.3\times 10^{-32}
(T/300\:{\rm K})^{-0.38}\:{\rm cm^6\:s^{-1}}$ for $T<300\:{\rm K}$
(Orel 1987) and $k_{\rm 3b}=1.3\times 10^{-32} (T/300\:{\rm
  K})^{-1}\:{\rm cm^6\:s^{-1}}$ for $T>300\:{\rm
  K}$ (Palla et al. 1983). The characteristic time for
the gas to become molecular is then
\beq
t_{{\rm H_2}}=\frac{1}{2k_{\rm 3b}\nh^2}=1.22\times 10^{16} K'
        \left(\frac{10^4\cm^{-3}}{\nh}\right)^{1.9}~~~~\yr.
\eeq
Equating this to the star formation time $t_*$ (which may be inferred
from equations \ref{eq:tstar} and \ref{eq:M} with $m_*=\esd
M/(1+f_d)$), we find that the gas becomes molecular at a density
$\nh=2.8\times 10^{11} K'^{5/7}~\cm^{-3}$, corresponding to
a mass $m_{*d}=1.34\esd K'^{5/4}~M_\odot$.  Omukai \& Nishi's (1998) and ABN's results show that
the gas becomes molecular at $T\sim 600$~K, so it has a slightly
different entropy parameter ($K'\simeq 0.4$) than the gas
at $\nh=10^4~\cm^{-3}$.  We conclude that the central
$0.4 M_\odot$ of the initial gas cloud is molecular, which
is close to the results of Omukain \& Nishi (1998) and ABN (as shown in the final 
timestep of their simulation).

\section{Rotating Infall}
\label{S:infall}

The collapsing clouds formed in the simulations of ABN and BCL are
rotating. When the initial angular momentum is very high, then the
structure may form a rotationally-supported disk on quite large scales
$\sim 10^3\sm$ (Bromm et al. 1999). It then fragments into 
protostellar cores with
masses $\sim 100\sm$. For more typical cosmological initial
conditions, ABN find that the first ``molecular cloud'' that forms in
their simulation does not fragment. It is rotating, but never (on the
scales resolved $\sim 20$~AU) fast enough to be completely 
supported by rotation.
Their figure 4b shows that the radial mass-weighted
rotational velocities are about half of Keplerian for a broad range of
the enclosed mass. These rotational velocities are therefore also
about the same magnitude as the sound speed.
ABN attribute the angular momentum transport needed to maintain
rotational velocities $\vrot\simeq \frac 12 \vkep$ to 
weak shocks in the flow. They were unable to continue their
calculation to the point that supersonic infall commenced.
Once such infall starts, it will be difficult to transfer
angular momentum from inside the sonic point to outside
(indeed, in the absence of shocks in the flow, it would
be impossible).  We idealize the angular momentum transport
in the infalling gas by assuming that $\vrot\propto\vkep$
outside the sonic point
(specific angular momentum $j\propto r^{1/2}$)
and that the angular
momentum is constant inside.
The angular momentum of the gas that accretes onto
the star-disk system is then characterized by the
parameter
\beq
f_{\rm Kep} \equiv \frac{\vrot(\rsp)}{v_{\rm Kep}(\rsp)} = 
       \frac{\vrot(\rsp)}
       {(GM_{\rm sp}/\rsp)^{1/2}},
\eeq
where $\rsp$ is the radius of the sonic point
and $M_{\rm sp}$ is the mass interior to it.  
Averaging over spherical shells, ABN found
$\vrot\simeq 0.5\vkep$, independent of $r$, so we adopt 0.5
as a fiducial value for $f_{\rm Kep}$.

%cfm Not clear why this is needed here:
%For the Hunter settling solution the sonic
%point starts moving outwards from the center at about the sound speed
%right after the protostar forms ($t=0$). 
%cfm Already mentioned above:
%Note that this phase was not
%simulated by ABN.  
%cfm Already said:
%Note that to reach this state there must be transport of angular momentum
%as the core is forming, and this is thought to be due to the turbulent
%motions in the gas.
%cf

%cfm This is trivial, so I have omitted it; furthermore, it
%    has nothing to do with angular momentum conservation.
%The outer radius, $r_d$, of the accretion disk that forms
%when angular momentum is conserved is given by
%\beq
%v_{\rm Kep}^2=\frac{G m_{*d}}{r_d},
%\eeq
%where $v_{\rm Kep}$ is the Keplerian velocity and 
%cf

%cfm Broke up array into two eqs. and inserted mass-radius reln
     Conservation of angular momentum from the sonic point
to the outer radius of the accretion disk, $r_d$, implies
\beq
r_d = f_{\rm Kep}^2 \left(\frac{M_{\rm sp}}{m_{*d}}\right)
    r_{\rm sp},
\label{eq:rd0}
\eeq
where the mass interior to $r_d$ is
$m_{*d}=(1+f_d)m_*$. 
%jct 
We shall typically assume that no mass is diverted from the inflow
between $r_{\rm sp}$ and $r_d$, so that in the above equation, $M_{\rm
  sp}=m_{*d}$.
%jt
In order to evaluate 
%jct
$M_{\rm sp}(r_{\rm sp})$, 
%jt
we note that the
mass-radius relation of the equilibrium core is
\begin{eqnarray}
M & = & \left[ \left(\frac{4\pi}{3-k_\rho}\right)^{1-\gamma_p} 
  \left(\frac{k_P K}{G}\right) r^{4-3\gamma_p}\right]^{1/(2-\gamma_p)},
  \nonumber\\
 &\rightarrow & 
 980 \left(\frac{r}{\rm pc}\right)^{7/9}
 K'^{10/9}\sm
\label{eq:mr}
\end{eqnarray}
from equation (\ref{eq:rho}) together with the relation
$M=4\pi r^3\rho/(3-\krho)$.
The mass inside the sonic point is made up of
the core of mass $m_0 c_{\rm th}^3 t/G$ (with $m_0=2.577$ 
for the isothermal settling solution vs. $m_0=0.975$ for the
Shu solution) and the infall
envelope, which has a mass $\simeq c_{\rm th}^3t/G$ for both solutions
(see Fig. 3 in Hunter 1977; most of the mass is near $\zeta\sim 1$). 
Thus in the isothermal case the mass within the
sonic point radius of the settling solution is about
$(2.577+1)/(0.975+1)\simeq 1.8$ 
times greater than in the equilibrium state. We assume a
similar increase applies to the $\gamma_p\simeq 1.1$ case. 
%cfm THE FOLLOWING IS BAD: IF WE ARE GOING TO KEEP \esd AS A PARAMETER IN
%    OUR THEORY, WE SHOULD DO IT EVERYWHERE.  SHOULDN'T WE SAY THAT
%    M_sp IS SMALLER THAN M BY A FACTOR \esd?
%jct removing:
%%The above
%%analysis assumes a quasi-spherical core, which is able to collapse with
%%100\% efficiency ($\esd=1$). 
If the protostar generates a
bipolar outflow that sweeps up material, then the mass inside 
the sonic point will be smaller by a factor of about $\esd$.
Using this as the efficiency at this stage,
and under the assumption that no mass is diverted from the inflow between
$r_{\rm sp}$ and $r_d$, we then have 
$m_{*d}\simeq M_{\rm sp} = 1.8 \esd M(r_{\rm sp})$.
Numerical evaluation of equation
(\ref{eq:rd0}) then yields    
\beq
r_d=3.44\left(\frac{f_{\rm Kep}}{0.5}\right)^2 \esd^{-9/7}
    \left(\frac{m_{*d}}{\sm}\right)^{9/7}
    K'^{-10/7}\:{\rm AU}.
\label{eq:rd}
\eeq
%cf
%where we have used equation (\ref{eq:mr}), boosted by a factor 1.8, 
%to replace
%$r_{\rm sp}$.
%jt (this implicitly assumes that $\esd\simeq1$).

\begin{figure}[h]
\begin{center}
\epsfig{
%rotenv
        file=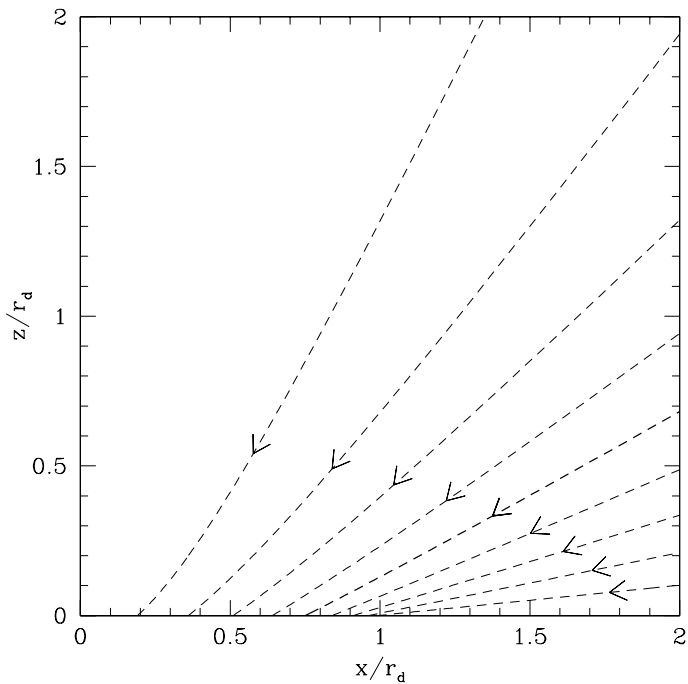,
        angle=0,
        width=\figwidth}
\end{center}
\caption{
\label{fig:rotenv}
Streamlines of the rotating, infalling envelope, projected onto the
meridional plane. The protostar is at (0,0). The regions between each
streamline and/or the axes, when rotated about the $z$-axis,  each
contain 10\% of the total accretion flow from this hemisphere. }
\end{figure}

Equation (\ref{eq:rd}) defines the outer radius of the disk. Material
falls on to this disk at all radii $r_*<r<r_d$ (Figure
\ref{fig:rotenv}). There is also some direct accretion to the star of
material that had very little initial angular momentum. We follow
Ulrich (1976) in describing the density distribution of this
freely-falling and rotating accretion envelope.  
Note that Ulrich and those who followed 
(Cassen \& Moosman 1981;
Terebey, Shu, \& Cassen 1984) assumed that the
rotation was initially uniform, so applying this solution to
a turbulent, differentially rotating cloud is necessarily 
approximate.
The fact that the
characteristic disk scale is generally much greater than the stellar radius
(which is always less than a few hundred solar radii, \S\ref{S:evol})
implies that most accretion proceeds via a disk. 
Provided that $r_d\geq r_*$, 
the mass accreting directly onto the disk is
\beq
\dot m_{\rm disk,\; direct}=\mdsd
     \left(1-\frac{r_*}{r_d}\right)^{1/2},
\eeq
whereas that accreting directly onto the star is
\beq
\dot m_{*,\;\rm direct}=\mdsd\left[1-\left(1-\frac{r_*}{r_d}\right)^{1/2}
     \right]
\eeq
(Adams \& Shu 1986).  Once the disk is large compared to the
star, direct accretion onto the star accounts for only a fraction
$r_*/2r_d$ of the total.

The properties of protostellar disks supplied by inflow with the
geometry shown in Figure \ref{fig:rotenv} have been considered by
Cassen \& Moosman (1981). The impact of gas streamlines with the disk
provides some viscosity to enable accretion through the disk, although
this is not important for $r\ll r_d$. 
%cfm DOESN'T THIS MATERIAL REPEAT WHAT WE HAVE ALREADY SAID? 
%    HAVEN'T WE SAID THAT WE ARE NOT INCLUDING MAGNETIC FIELDS HERE?
%jct Since we just showed that most mass goes into a disk, I wanted
%to briefly discuss how this material may then reach the star
However, if the disk becomes
sufficiently massive we expect the gravitational turbulent viscosity
of clumps in a massive disk (Gammie 2001) or large scale ($m$=1 mode)
instabilities (Adams et al. 1989; Shu et al. 1990) to be the most
important processes for driving inflow.  Dynamo amplification of seed
magnetic field (Tan \& Blackman 2004) may lead to the development of
magneto-rotational instability (MRI) (Balbus \& Hawley 1991), which
can provide a source of viscosity. Fragmentation of a gravitationally
unstable disk may occur if the local thermal time of the disk becomes
less than about half the orbital time (Gammie 2001), but this does not
seem to occur in these disks (Tan \& Blackman 2004). The structure of
the inner accretion disk is considered in \S\ref{S:disk}.

%% We have modeled the structure of the disk using the theory of thin,
%% viscous disks (Shakura \& Sunyaev 1973), for various values of the
%% viscosity parameter $\alpha_{\rm ss}$, including $\alpha_{\rm
%%   ss}=0.3$, as may be appropriate for self-gravitating, clumpy disks
%% (Gammie 2001) and $\alpha_{\rm ss}=0.01$, which may be applicable to
%% disks in which the magneto-rotational instability is operating (Balbus
%% \& Hawley 1998). The energy equation is modified to include the energy
%% needed to ionize H and He, which is important in the earliest stages.
%% The details of these results are presented in a separate paper
%% (Tan \& Blackman 2003), where they are used to assess the possibility
%% of magnetic field amplification by a disk dynamo. For the purposes of
%% the current discussion, we assume that mass can be transported through
%% the outer accretion disk by viscous processes or large scale
%% gravitational instabilities. Our modeling of the disks indicates that
%% self-gravity is important during the early stages and for the outer
%% parts of the disk. The inner regions are not self-gravitating: for
%% example the Toomre stability parameter, $Q$, is greater than unity
%% inside radii of about 15 $r_*\simeq 100\:{\rm AU}$, for the fiducial
%% case when $m_*=10\sm$. The results of Tan \& Blackman (2003) indicate
%% that magnetic fields may be generated in the disk, so we set
%% $\alpha_{\rm ss}=0.01$ for all the models of the inner disks presented
%% here.

\begin{figure}[h]
\begin{center}
\epsfig{
%rotenvden
        file=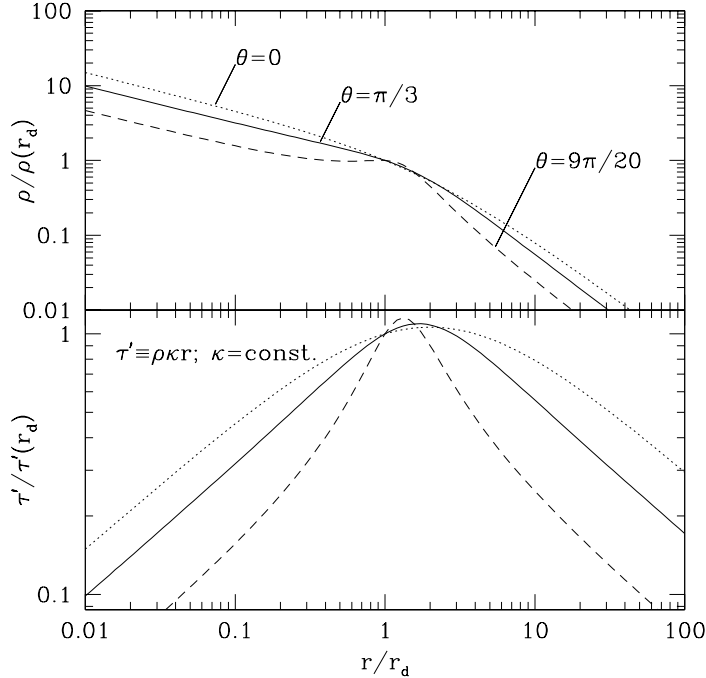,
        angle=0,
        width=\figwidth}
\end{center}
\caption{
\label{fig:rotenvden}
{\it Top panel:} Density of flow relative to that at $r_d$ along radii at
angles 0 ({\it dotted}), $\pi/3$ ({\it solid}), and $9\pi/20$
({\it dashed}) from the rotation axis. Note that the absolute
densities at $r_d$ of these three cases are in the ratios
0.33:0.59:1.43 relative to the spherical case. {\it Bottom panel:}
Characteristic optical depth ($\tau^\prime$) in traversing length
scale $r$ (assuming constant opacity), relative to that at $r_d$, for the
same cases as above. Although $\kappa$ is a more complicated
function of $\rho$ and $T$, this figure illustrates that the optical
depth near the star is much reduced with respect to the spherical case
and that under certain conditions when the envelope is already
optically thick, it may become optically thin from the inside out, as the
overall mass accretion rate declines.}
\end{figure}

One consequence of the geometry of rotating infall for initial
protostellar core formation is that the optically thick stages of the
collapse (with respect to both molecular line and continuum cooling,
see \S\ref{S:structureform}) are reached at somewhat higher masses
than in the case of spherical accretion.  Since material can dissipate
some of its energy in the disk, it has a smaller bulk kinetic energy
when it reaches the star.  It also has a higher temperature because of
the dissipation of this energy, which has implications for the
temperature of the post-accretion shock relaxation region and the
protostellar evolution (see below). By changing the size and
temperature of the photosphere around the star, disk accretion tends
to create a hotter radiation field, which has an important effect on
the feedback processes that occur (\S\ref{S:lum}; Paper II). The
density and a simple model of the optical depth near the star are
shown in figure \ref{fig:rotenvden}.

\section{Properties of the Inner Accretion Disk}
\label{S:disk}

Since much of the accretion onto the star goes through an accretion
disk, it is necessary to determine the properties of this disk,
particularly near the star where the disk emission can have
significant feedback effects.  Our objective in this section is to
estimate the luminosity and approximate spectrum emitted from the
inner region of the accretion disk. The luminosity from the inner part
of the accretion disk can be comparable to that produced at the
boundary layer. The structure of the accretion disk, and in particular
its scale-height, is also of interest because it affects the size of
the region where the accretion luminosity from the boundary layer is
released, and thus the spectrum.  The disk may be the site of dynamo
action that can amplify weak seed magnetic fields and perhaps generate
a hydromagnetic bipolar outflow, a common occurrence in contemporary
star formation. For the primordial case this question has been
addressed by Tan \& Blackman (2004), where it is concluded that strong
large-scale magnetic fields are likely to be produced if turbulence in
the disk generates helicity, particularly by the time the protostar is
massive and contracting to the main sequence. Even in the absence of
magnetic fields the disk may generate a radiatively driven outflow
(Paper II), which depends somewhat on the structure of the disk.
Finally the disk may act to shadow gas in the equatorial plane from
the brunt of the stellar radiative feedback, thus influencing the
efficiency of star formation.

In order for the star to continue accreting, it must keep its rotation
rate slower than the break-up rate. This is most likely achieved
either by the action of magnetic fields coupled to the larger scale
disk or an outflow, or by the excitation of density waves in the disk
by the star (if it is rotating so quickly that it starts to assume a
non-axisymmetric structure). We have assumed that these processes
occur; furthermore, we assume that they are sufficiently effective
that they maintain the overall rotation of the star to be much less
than break-up, so that the rotational energy is negligible.

To calculate the radial structure of the inner accretion disk at any
given point in the evolution of the protostar, we assume that it is
fed smoothly at a rate given by equation (\ref{eq:mds3}) and use the
standard theory of steady, thin, viscous accretion disks, with a
spatially constant viscosity parameter, $\alpha_{\rm ss}$ (Shakura \&
Sunyaev 1973; Frank, King, \& Raine 1995). 
The viscosity is assumed to be a function of the total
pressure, which is the sum of gas and radiation pressures. We have
evaluated two cases with $\alpha_{\rm ss}=0.01,0.3$ but regard the
$\alpha_{\rm ss}=0.01$ case as the most appropriate for the very inner regions
that are relevant here (see Tan \& Blackman 2004 for a discussion on
these choices and for the full results from both cases). This is a
typical value for viscosity generated by the magneto-rotational
instability (MRI) (Balbus \& Hawley 1991, 1998), although the
dispersion in quoted results from numerical studies is about two
orders of magnitude. It must be stressed that the use of a spatially
constant ``$\alpha_{\rm ss}$'' viscosity model is motivated by its theoretical
simplicity and convenience, and results of this modeling should be
viewed as being only an approximate guide to reality.

The inner scale of the disk is set by $r_*$, which we estimate in
\S\ref{S:evol} below. We ignore the effects of energy injection from
the star, deferring this to paper II. We again use the opacities of
zero metallicity gas of Rogers \& Iglesias (1992) and Iglesias \&
Rogers (1996) for $T\ga 6000 \: {\rm K}$ and of Lenzuni et al. (1991)
for $T\la 6000\: {\rm K}$ (however, here the interest is only in the
higher temperature regime).  The method of solution assumes that the
mean disk temperature, $\bar{T}$, is much greater than the surface
temperature, $T_{\rm eff}$, which is a good assumption for the inner
regions that we consider, except in the early stages.

The high accretion rates and large sizes of primordial protostars cause the
ionization energy to have an important effect on the disk structure, 
particularly during the earlier stages of the evolution. 
The energy equation for the disk is worked out in Appendix
A.1: 
\beq
F= \frac{3 G m_*\mds}{8 \pi r^3}
         \left[1-\left(\frac{r_*}{r}\right)^{1/2}\right]
         + \frac{\mds}{4 \pi r} \frac{d}{dr}\left(\frac 53 \bar\epsth
        + \bar\epsilon_I \right),
\label{eq:diskenergyion}
\eeq 
where $\bar\epsth$ and $\bar\epsilon_I$ are 
the thermal energy and dissociation/ionization energy per unit mass, averaged
over the thickness of the disk. The first term on the right
hand side of eq.~(\ref{eq:diskenergyion}) is the viscous dissipation per
unit disk face area. 
Note that usually $d\bar\epsth/dr<0$ and
$d\bar\epsilon_I/dr<0$. 
The effect of the
thermal energy term is small for thin disks such
as those  we consider, so we drop
it from our numerical calculations.
However, the dissociation/ionization term, which
is often neglected, can be an order
of magnitude larger than thermal term and can have a
major impact on the disk structure. We therefore retain this
term.

\begin{figure}[h]
\begin{center}
\epsfig{
%diskexam01paper1
        file=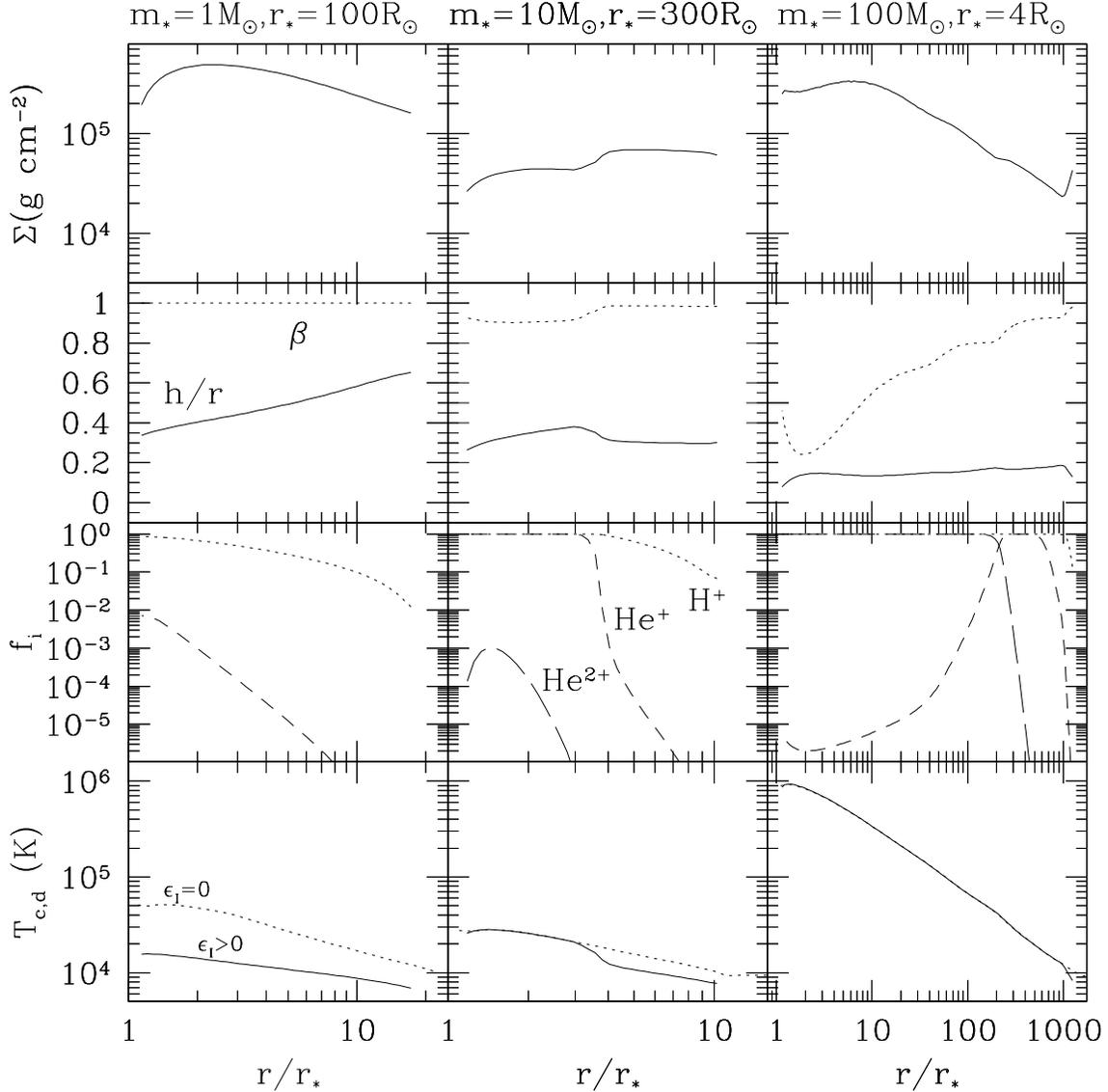,
        angle=0,
        width=6.2in}
\end{center}
\caption{
\label{fig:diskexam01}
Protostellar disk structure for models with $\alpha_{ss}=0.01$ and
$m_*=1,10,100\sm$, for which $r_*=100,300,4R_\odot$ (see
\S\ref{S:evol}) and $\mds=(17,6.4,2.4)\times 10^{-3}\smyr$ (see
\S\ref{S:accretion}),
respectively.  From top to bottom the panels show (1) surface density,
$\Sigma$; (2) ratio of scale-height to radius, $h/r$, and ratio of gas
pressure to total pressure, $\beta$; (3) ionization fractions of $\rm
H,\:He^+,\:He^{2+}$; and (4) central disk temperature, $T_{\rm c,d}$ (the dotted lines show
results for when the ionization energy is neglected).  }
\end{figure}

In this paper, we give an approximate solution for the disk
based on the treatment of Frank et al. (1995).  They
integrate the radiative diffusion equation to obtain
$F\simeq 4\sigma T_{\rm c,d}^4/3\tau$, where
$T_{\rm c,d}$ is the midplane temperature and $\tau$ is the optical
depth from the midplane to the surface.  
To solve the structure of the disk we divide it into a large number of
radial zones, and start at the outer edge with an assumed ionization
state that is almost completely neutral. We then solve the modified
disk structure equations to find the temperature, density, ionization
state of H and He (from the Saha equation), etc., and 
then advance the solution 
inwards given the new ionization state.
Figure~\ref{fig:diskexam01} shows three examples of
disk structure for $\alpha_{ss}=0.01$, with $m_*=1,10,100\sm$. At
these masses the stellar sizes are taken to be
$r_*=100,300,4\:R_\odot$ and the accretion rates 
in the inner disk---i.e., onto the star---are
$\mds=(17,6.4,2.4)\times 10^{-3}\smyr$, respectively (eq. \ref{eq:mds3}).

As we shall see below, at early times the large accretion rates in
primordial protostars lead to large rates of energy absorption by
ionization and dissociation (eq. \ref{eq:LI}), and this can regulate
the midplane temperature to about $10^4\:{\rm K}$ in the early stages.
The effect of including and not including the ionization energy is
shown in the temperature plots of Figure~\ref{fig:diskexam01}.

Our disk model breaks down when the predicted surface temperature is
very cool ($T\la 5000\:{\rm K}$), which occurs for low stellar masses.
This break down occurs because in our thin disk model we approximate
the opacity by its value at the midplane, whereas in fact the opacity
increases rapidly with temperature for $T\la 10^4$~K.  
%jct - removed this part since we simplified fig 4.
%For example,
%the 1 $M_\odot$ model in Figure \ref{fig:diskexam01} suggests that the
%surface temperature of the inner disk might be as low as a few hundred
%degrees because most of the energy has gone into ionizing the
%accreting material.  In this model, the central temperature of the
%disk is about $10^4$ K, so that (1) the opacity is high and (2) the
%gas is partially ionized.  Had we resolved the vertical structure of
%the disk, then the average opacity and hence the central temperature
%would have been lower; with less ionization, the effective temperature
%would have been higher. 
However, since we are most interested in the
radiative feedback from the star and disk at higher masses, we do not
attempt to correct this limitation.% of the thin disk model.
%jt

\section{Protostellar Evolution to the Main Sequence}
\label{S:evol}

The basic process we wish to model is the evolution of the protostar's
radius as it grows in mass by accretion of material from an accretion
disk or directly from the infalling envelope. Stahler et al. (1986)
and Omukai \& Palla (2001) considered a detailed model for the
evolution of a protostar with spherically symmetric accretion and a
constant accretion rate of $4.4\times 10^{-3}\smyr$. Like these
authors, we also define the protostellar radius to be the location of
the accretion shock (see Appendix \ref{S:direct}). 
This size is important for determining the
accretion luminosity of the protostar and it sets a lower limit for
the photospheric radius of the protostar, which controls the spectrum
of emitted radiation.

In order to be able to follow the evolution of a protostar for an
arbitrary accretion rate, we use the analytic approach developed by
Nakano et al. (1995; 2000).  They estimated the protostellar radius by
equating the rate of accretion of energy by the protostar (which is
worked out in Appendix A) with the rate of change of the energy of the
protostar,
\beq
E=-\frac{ a_g\beta}{2}\;\frac{Gm_*^2}{r_*}+m_*\bar\epsilon_{I}.
\label{eq:E}
\eeq
The first term in this equation represents the sum of the thermal
energy of the gas, the energy of the radiation and the gravitational energy,
and it follows directly from the virial theorem.  
Here $\beta$ is the mean ratio of gas pressure to total pressure and
$a_g Gm_*^2/r_*$ is the gravitational energy.
We approximate the structure of the protostar as a polytrope of index
$n$, so that
$a_g=3/(5-n)$ with $n<5$.
The second term in equation (\ref{eq:E}) is the mean ionization
energy of the material in the star.  We shall assume that
the star is approximately
fully ionized ($\bar\epsilon_I\simeq \epsilon_{Im}=16.8\;{\rm
eV}/m_{\rm H}$), which
means that we cannot treat the earliest stages of protostellar evolution
with our method.
Differentiating equation (\ref{eq:E}) 
and equating it to the rate of energy accretion accretion
given in equation (\ref{eq:dedt}) evaluated
just inside the postshock relaxation layer of the accretion
shock (denoted $r_2$)
yields
\beq
\frac{d\ln r_*}{d\ln m_*}=2+\frac{d\ln\beta}{d\ln m_*}
            -\frac{4}{a_g\beta \vffs^2}\left[
           \frac 12 \vffs^2+\epsilon_{Im} -\left\langle\frac{5kT_2}{2\mu_2}
           +\epsilon_{I2}\right\rangle+\frac{1}{\mds}\left(
           L_2-\enucd\right)\right],
\label{eq:drdm}
\eeq
where $\mu_2$ is the mean mass per particle at $r_2$
and $\enucd$ is the rate at which nuclear reactions release
energy in the protostar.
This result is identical to that
of Nakano et al. (2000), except that (1) equation (\ref{eq:drdm})
has the enthalpy $5kT/2\mu$ instead of the thermal 
energy $3kT/2\mu$, (2) we allow the accretion flow to
be optically thick, so that it is possible for $T_2$ to be
much larger than the photospheric temperature, and
(3) we explicitly allow for both direct accretion onto
the star and for accretion via a disk.

        We now consider the evaluation of each of the terms
in this equation.  First, the term $d\ln\beta/d\ln m_*$
becomes significant only when radiation pressure becomes important,
which occurs only at
relatively high masses. The appropriate polytropic index in this regime
is $n=3$, and in this case the mean value of $\beta$ is the same as the
central one. We calculate $\beta$ assuming a standard
Eddington stellar structure model with $n=3$ applies to the protostar.

     Next consider the temperature behind the accretion
shock, which is discussed in Appendix
\ref{S:direct}. For accretion directly onto the star, the value
of $T_2$ in the optically thin case is given by equation
(\ref{eq:t2}). As remarked in Appnedix \ref{S:optthin}, we
allow for the non-blackbody nature of the radiation
field there by using the Saha equation to evaluate $\epsilon_{I2}$ at
$T_{\rm eff,\,2}$.
When the accretion is optically thick, we integrate through
the radiative precursor to find $T_2$, as described in Appendix
\ref{S:optthk}. In the optically thick case, $\epsilon_{I2}$
is evaluated at $T_2$.

        Disk accretion presents a more complicated problem.
Gas reaches the surface of the star at a temperature $\bar{T}$
that we have estimated from standard thin disk theory
(\S \ref{S:disk}).  However, as this gas joins the star, it
will spread over the stellar surface, where it can lose
heat. Integrating the radiative transfer equation through
the disk shows that the 
mean temperature is related
to the effective temperature by $\bar{T}\propto \tau^{1/4}T_{\rm eff}$
(Shakura \&  Sunyaev 1973).  For a constant opacity per unit
mass, $\tau\propto 1/A$, where $A$ is the area normal
to the direction of propagation of the radiation; similarly,
$T_{\rm eff}\propto 1/A^{1/4}$, so that altogether
$T\propto 1/A^{1/2}$.  For the disk, we identify the
$A$ as the area of the star covered by the disk,
$4\pi r_*^2\cos\theta_d$, where $\theta_d$ is
the angle from the axis at which the disk intersects
the star.  We therefore adopt the approximation that
for disk accretion, $T_2=\bar{T}(r_*)(\cos\theta_d)^{1/2}$.
In this paper, we estimate the height of the disk as
1.5 times the scaleheight; a more accurate estimate
will be given in Paper II.

To evaluate the luminosity $L_2$, which is effectively the luminosity
that can be transported radiatively, we follow Nakano et al. (1995)
and approximate it as being equal to the value of the ZAMS luminosity
(Schaerer 2002) of a star of the same mass. However, this is not a
very good approximation at lower stellar masses, $m_*\lesssim 10\sm$,
when, because of its expanded size, the protostar's internal
luminosity is much smaller than the ZAMS value (Stahler et al. 1986).
In this regime we evaluate $L_2$ from analytic fits to the results of
Omukai \& Palla (2003, fig. 5), so that $L_2\simeq 0$ for $m_*\lesssim
7\sm$, then rises as $L_2/L_\odot\simeq 390 (m_*/\sm)^2 - 2700
m_*/\sm$, which is used until $L_2\simeq L_{\rm ZAMS}$.  The idea
behind this approach of specifying $L_2(m_*)$ is that the amount of
radiative luminosity that a star can carry is determined by its
structure.  In a main sequence star, the rate of energy generation
adjusts to give this luminosity; in a rapidly accreting protostar, the
rate of energy generation, both nuclear and gravitational, can differ
substantially from that which can be transported, and this difference
can cause the star to either swell or shrink.
The evolution during the early stages is not particularly
sensitive to this approximation because the accretion luminosity is
much greater than the internal luminosity.

The rate of nuclear energy generation, $\enucd$, is determined by both
deuterium burning and the nuclear reactions that occur on the main
sequence.  We do not explicitly include the latter, but note that they
prevent $r_*$ from shrinking below its main sequence value.  Deuterium
burning sets in when the central temperature $T_{c,*}\simeq 10^{6}\:{\rm K}$.
The central temperature is 
\beq
T_{c,*}=\beta_{c} a_T\frac{\mu m_{\rm H}}{k}\frac{Gm_*}{r_*}
\label{eq:Tc}
\eeq where $\mu$ is the mean particle weight in the ionized interior and
$a_T$ is a quantity of order unity (e.g.  $a_T=0.54,0.84$
%$a_T=0.5385,0.838$
for $n=1.5,3$, Chandrasekhar 1939). For $m_*\la 10\sm$, which is the
regime where D-burning can be relatively important, $\beta_{\rm gas}$
and $\beta_{c}$ are close to unity.  The reaction rate rises sharply
with temperature so that, while there is fuel available, this process
can act as a thermostat and maintain an almost constant $T_{c,*}$ (Stahler
1988). If the protostar reaches this central temperature at relatively
low masses (as in contemporary star formation), then the luminosity
produced by D burning can be large compared to that which can be
transported radiatively through the star. The star becomes convective,
which allows freshly accreted deuterium to be brought to the center
for burning (we take ${\rm D/H}=2.2\times 10^{-5}$; Pettini \& Bowen
2001). The rate of D--burning can be calculated in this case using the
model of Nakano et al. (2000) (see also McKee \& Tan 2003). However,
for realistic accretion rates (i.e. from cores with $K'\sim 1$), the
protostar reaches the D--burning temperature only at relatively high
masses ($m_*\sim 10\sm$) so that the star remains radiatively stable,
D is quickly depleted from the stellar center, and the protostar can
continue to evolve to hotter central temperatures, little affected by
this nuclear energy generation. In this case we set
$\enucd=10^4\:L_\odot$, independent of stellar mass, which
approximates the more detailed calculations of Omukai \& Palla (2003).
This choice has only a minor effect on the stellar radius for most of the 
D-burning regime since usually $\enucd\ll L_2$, the 
internal protostellar luminosity 
(Fig. \ref{fig:Lstarevolpalla3testing}).
Eventually the central temperature becomes hot enough to allow support
via hydrogen burning. Some H--burning reactions start generating
significant luminosity when $T_{c,*}\gtrsim 2\times 10^7\:{\rm K}$ (we set
$\enucd=10^5\:L_\odot$ at this point, Omukai \& Palla 2003). However,
full support of the star only occurs once $T_{c,*}\sim 10^8\:{\rm K}$,
which is hotter than in the contemporary case, leading to smaller main
sequence radii (e.g.  Schaerer 2002). In this way the protostar
settles on the zero age main sequence, where it may continue to grow
in mass if the accretion is on-going.

Finally, the evolution of the polytropic index of the star needs to be
accounted for.  We take $n$ to be constant during specific phases of
the evolution, but allow for certain transitions as follows.  By
comparison to the results of Stahler et al. (1986), we find that an
initial value of $a_g=1.1$ (corresponding to $n=2.3$) is a reasonable
description of the initial structure, while the star is younger than
its Kelvin-Helmholz time.  If the central temperature becomes hot enough for
D-burning and $\enucd\gtrsim L_2$, so that
the star becomes convective, then $n=1.5$ ($a_g=0.86$) (this case is
not realized in our models with $K'=1$).  In this case, the star
eventually becomes radiative ($n=3$ and $a_g=1.5$). The star also
relaxes to this structure in the absence of convective D-burning,
after the age of the star becomes greater than the current Kelvin-Helmholz time
(this is the case realized in all the models presented in this paper).
The expansion of the protostar after a Kelvin-Helmholz time reflects the fact
that it is relaxing to a globally more compact state, which releases
gravitational energy and causes the outer layers of the star to
expand. From comparison with the results of Omukai and Palla (2001) we
set this expansion factor to be three. For protostars that evolve from
a convective D-core burning state to a radiative state including
D-shell burning, there is an analogous expansion of the outer layers,
by about a factor of two (Palla \& Stahler 1991).

We take our initial condition to be a protostar of mass $m_*=0.3\sm$
and radius $r_*=30\:R_\odot$, which is a reasonable extrapolation of
the numerical results of Ripamonti et al. (2002), who formed a core
with a mass of $0.04\sm$ and a size of $10^{12}\:{\rm cm}\simeq
14\:R_{\odot}$ at the end of the core formation phase of their simulation (see also Omukai \&
Nishi 1998). In any case, since the gas in front of the accretion
shock is optically thick at early times, the subsequent evolution is
insensitive to the initial condition.  

A summary of the basic features of the protostellar evolution is as
follows. The mean density and central temperature ($T_{c,*}\sim 10^5\:{\rm
  K}$) of the initial state are relatively low compared to typical
stellar values. However, as the mass increases via accretion,
self-gravity compresses the star and raises the central temperature.
The rate of this contraction is limited by the Kelvin-Helmholz time;
indeed rapid accretion of material on shorter timescales swells the
star. The star is typically able to reach a mass $\sim 10\sm$ by the
point that its age is about equal to its instantaneous Kelvin-Helmholz time.
It also reaches the D--burning temperature at about this time: too
high a mass for D--burning to create a convective core.  Structural
rearrangement to a radiative ($n=3$) core temporarily expands the
outer layers of the star by factors of a few. However, the rapidly
increasing internal luminosity causes fast contraction (see eq.
\ref{eq:drdm}) that is only halted once the star reaches the zero age main
sequence, where the central temperature ($T_{c,*}\sim 10^8\:{\rm K}$) is
hot enough for H--burning reactions to support the star. This occurs at
stellar masses of about $100\sm$. In our
numerical implementation of the above evolution we allow for accretion
from both a disk and directly to a star, the relative proportions of
which depend on $f_{\rm Kep}$, the size of the star, and the mass that
has collapsed. 

Observe that because of the high accretion rates the total energy
absorbed by dissociation and ionization (and the corresponding reduction
in the protostar's luminosity that escapes to infinity) can be quite
large:
\begin{eqnarray}
L_{\rm I,max}\equiv && \left( \frac{dE}{dt}\right)_{\rm I,max}
       =\mds\epsilon_{Im}\\
       &=& 2.65\times 10^{3}\left(\frac{\mds}{10^{-2}\smyr}\right)\:L_\odot.  
\label{eq:LI}
\end{eqnarray}
This process helps to keep the radiated luminosity sub-Eddington
during the early stages of protostellar evolution. In the later
evolution of models of protostars accreting from realistic rotating
cores, then equation (\ref{eq:LI}) applies to the reduction in the
accretion disk luminosity, as discussed in \S\ref{S:disk}.

\begin{figure}[h]
\begin{center}
\epsfig{
%Rstarevolpallanewfinal
        file=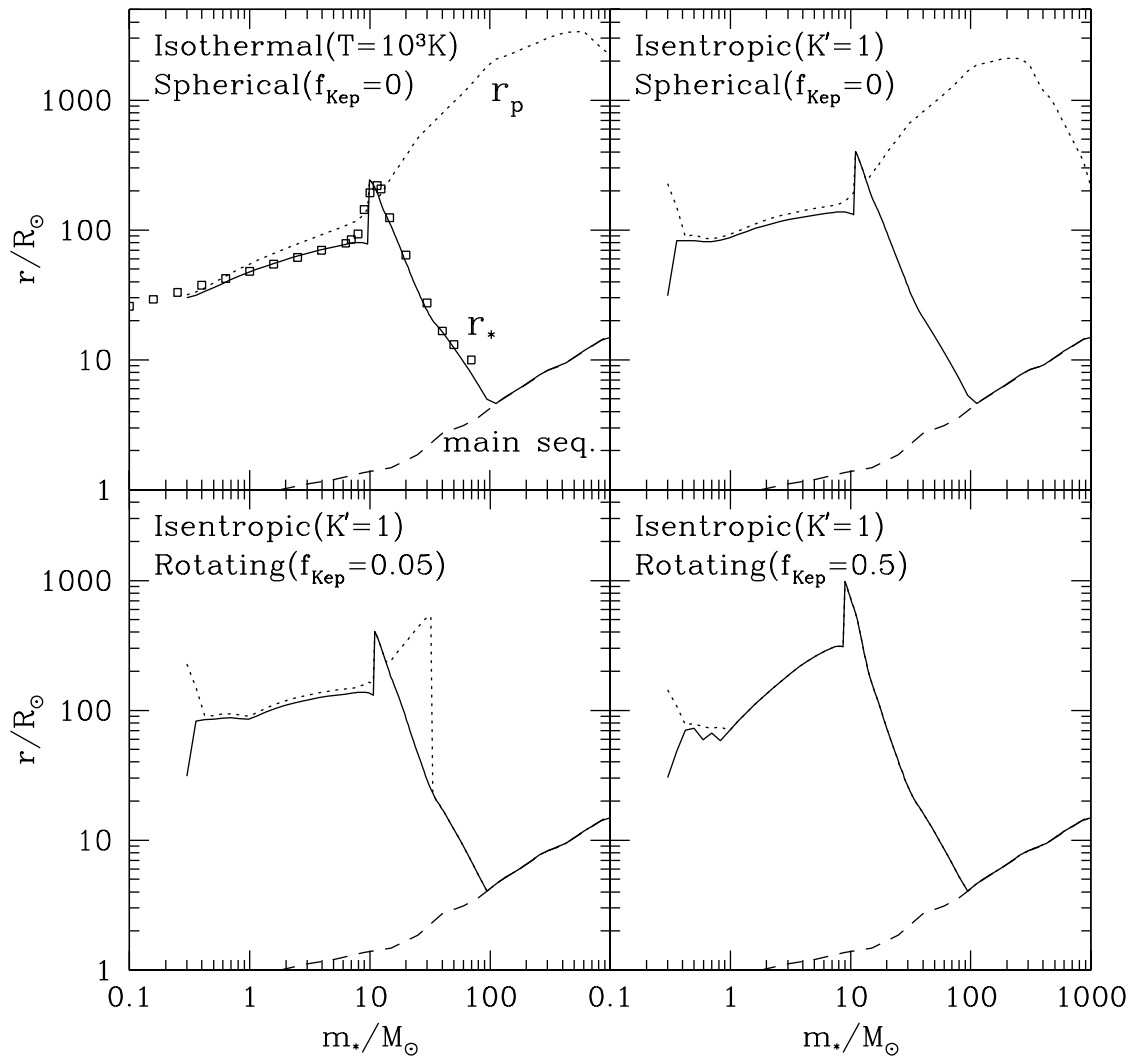,
        angle=0,
        width=6in}
\end{center}
\caption{
\label{fig:Rstarevolpallanew}
Protostellar (solid lines) and photospheric (dotted lines) radii as a
function of protostellar mass. The dashed line in each plot is the
zero age main sequence radius (Schaerer 2002). (a) Top left: Spherical
isothermal accretion at a rate $4.4\times 10^{-3}\smyr$ corresponding
to $T=1000\:{\rm K}$.  The squares are the results of Stahler et al.
(1986) and Palla \& Omukai (2001).  (b) Top right: Fiducial spherical
isentropic accretion model. (c) Bottom left: isentropic case with
small amount of rotation: $f_{\rm Kep}=0.05$. The abrupt decline of
the photosphere between 30 and 40~$\sm$ reflects the development of an
inner photosphere as the density declines (see Figure
\ref{fig:rotenvden}). (d) Bottom right: Fiducial isentropic case with
$f_{\rm Kep}=0.5$.  }
\end{figure}

A comparison of our simple semi-analytic model with the spherical accretion test case of
Stahler et al. (1986) and Omukai \& Palla (2001) is shown in Figure
\ref{fig:Rstarevolpallanew}a. As discussed above, the expansion factor
of the luminosity wave has been adjusted to give agreement with the
models.  We then applied the same model to the accretion rate of
equation (\ref{eq:mds3}), but still with spherical symmetry.  We show
two examples of rotation for the isentropic model: the fiducial
$f_{\rm Kep}=0.5$ case and one with $f_{\rm Kep}$ ten times smaller.

Rotation causes the infall envelope to become optically thin at much
smaller masses. This causes the photospheric radius to be much closer
to the stellar surface, leading to a hotter radiation field (\S\ref{S:lum}).
Also $f_k$ is reduced, because most
material is processed through an accretion disk and so there is less
bulk kinetic energy to advect, even if conditions are optically thick.
However, the conditions at point ``2'', just behind the accretion
shock are hotter in the disk case, so that overall more energy is
advected into the star and its size is bigger.

\section{Radiation from the Protostar}
\label{S:lum}

\begin{figure}[h]
\begin{center}
\epsfig{
%Lstarevolpalla3testing
        file=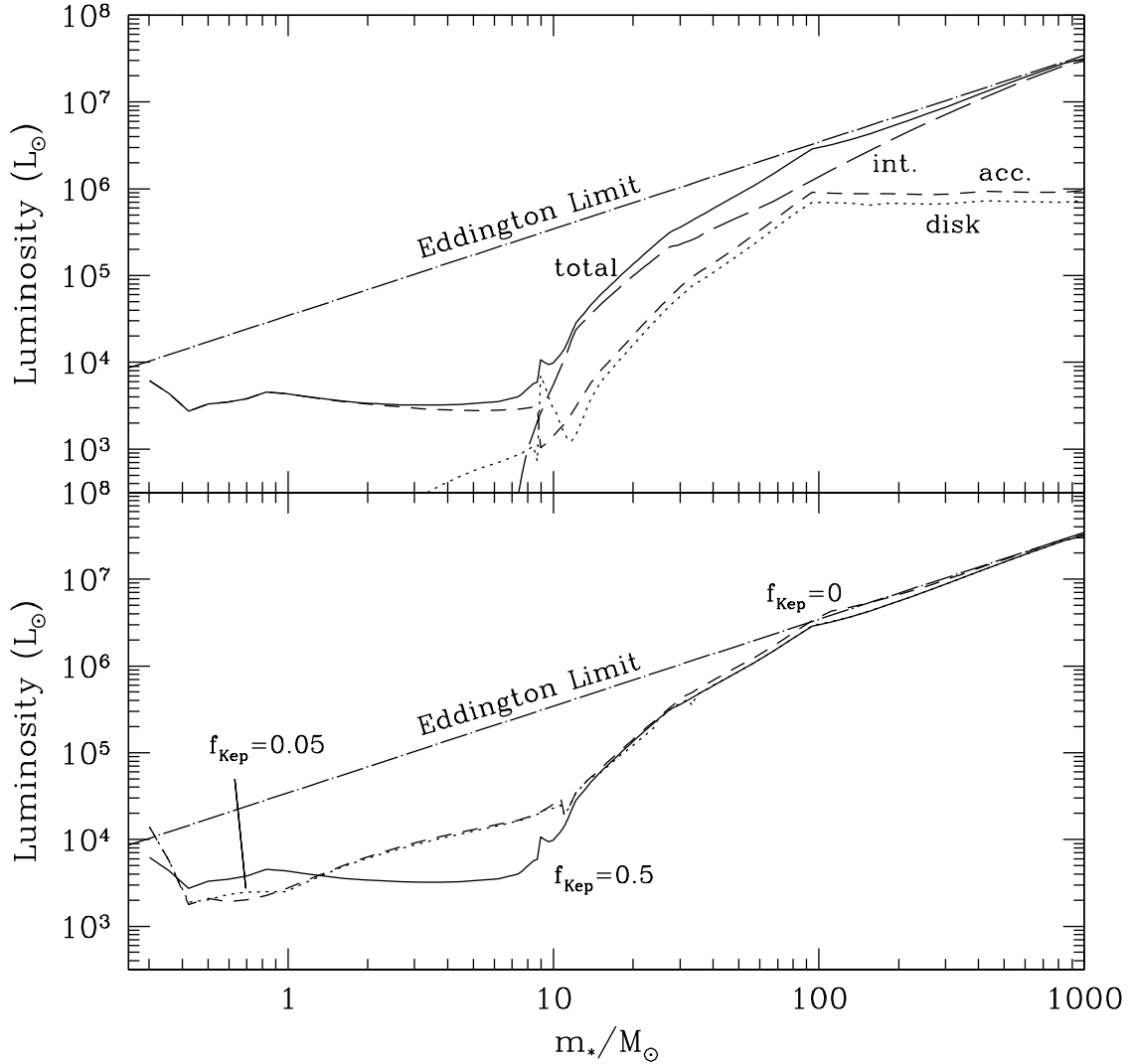,
        angle=0,
        width=6in}
\end{center}
\caption{
\label{fig:Lstarevolpalla3testing}
Top panel: Evolution of bolometric luminosities with protostellar mass
for the fiducial isentropic model with rotation $f_{\rm Kep}=0.5$,
$K'=1$, and $\alpha_{\rm ss}=0.01$.  Total radiated bolometric
luminosity (solid line) including contribution from inner accretion disk,
stellar accretion luminosity including from direct spherical
accretion and boundary layer accretion (dashed line), internal
protostellar luminosity (long-dashed line), and accretion disk
luminosity from $r<10r_*$ (dotted line) are shown. The Eddington
luminosity is shown by the dot-long-dashed line.  The combination of a
declining accretion rate, increasing stellar radius, and increasing
stellar mass at later stages when the protostar is accreting on the
main sequence, lead to approximately constant accretion luminosities.
Note the total luminosity remains sub-Eddington over the entire
evolution ($m_*\leq 1000 M_\odot$.  
Bottom panel: Evolution of total bolometric luminosities for fiducial
isentropic models with $K'=1$ and $\alpha_{\rm ss}=0.01$, and with
rotation $f_{\rm Kep}=0,0.05,0.5$ (dashed, dotted, solid lines, respectively).
Note that the spherical accretion case ($f_{\rm Kep}=0$) exceeds the Eddington limit
at $m_*\simeq 100\sm$, but the more realistic disk accretion models are slightly sub-Eddington.
}
\end{figure}

We are now in a position to calculate 
the total bolometric luminosity of the protostar. This luminosity
is the sum of the
radiation from the stellar surface, the boundary layer, and that from
the inner accretion disk; we somewhat arbitrarily include emission out
to a distance $10r_*$ in the latter. Emission from further out in the
disk is not significant for the feedback processes we consider in
Paper II.  The luminosity from the boundary layer is comparable to,
but almost always somewhat greater than, the luminosity from the inner
disk.  At high stellar masses the internal luminosity of the star
becomes dominant. Figure \ref{fig:Lstarevolpalla3testing}a shows the
evolution of the total luminosity and its various sub-components for
the fiducial $f_{\rm Kep}=0.5$, $K'=1$, $\alpha_{\rm ss}=0.01$ case.
The luminosites of the models with different rotational parameters are
quite similar (Fig. \ref{fig:Lstarevolpalla3testing}b). The main
differences are due to the different protostellar sizes (and thus
accretion luminosities) and to the fact that our definition of the
total luminosity only includes energy generated in the immediate
vicinity of the star ($r<10r_*$) and not, for example, the outer
accretion disk. This last point causes spherically-accreting
protostars to have a higher total luminosity than disk-accreting ones,
and is enough to make the difference between whether the star is sub-
or super-Eddington at around $100\sm$. This measure has been invoked
in the spherical case as an important factor in determining the final
mass of the star (Omukai \& Palla 2003). While we regard the question
of mass-limits due to feedback as being more complicated than a comparison
of the protostar's luminosity to the Eddington value (Tan \&
McKee 2003; Paper II), Figure \ref{fig:Lstarevolpalla3testing}b shows
that if such criteria are to be used, then the dependence on the
geometry of the accretion must be accounted for.

We calculate the spectra, and in particular the H--ionizing photon
luminosity, $S$, of the various components as follows.  When the
accretion flow is optically thin, the luminosity from the stellar
surface is $4\pi r_*^2\sigma T_{\rm eff,\, 1}^4$, where $T_{\rm eff,\,
  1}$ is given in equation (\ref{eq:teff1}).  When the accretion flow
is opaque, the effective temperature and radius of the photosphere are
determined as described in \S \ref{S:optthk}. The assumption of a
blackbody spectrum for the stellar spectrum is quite accurate for
calculation of fluxes of hydrogen-ionizing photons (Schaerer 2002),
though not for photons that ionize He. Fortunately it is the feedback
effects from H--ionization that are much more important (Paper II).
The contribution from the boundary layer accretion luminosity is
calculated assuming the radiation emerges with a blackbody spectrum
from the upper and lower surfaces of an annulus that extends radially
from the stellar surface by a distance equal to the height of the
disk's photosphere (estimated to be about 1.5 density scaleheights,
$h$, from the disk model of \S\ref{S:disk}; a more accurate treatment
is given in Paper II).  Finally the contribution from the inner
accretion disk ($r<10r_*$) is included, using the surface temperatures
predicted by the radial disk model (\S \ref{S:disk}). 

\begin{figure}[h]
\begin{center}
\epsfig{
%Sstarevolpalla3
        file=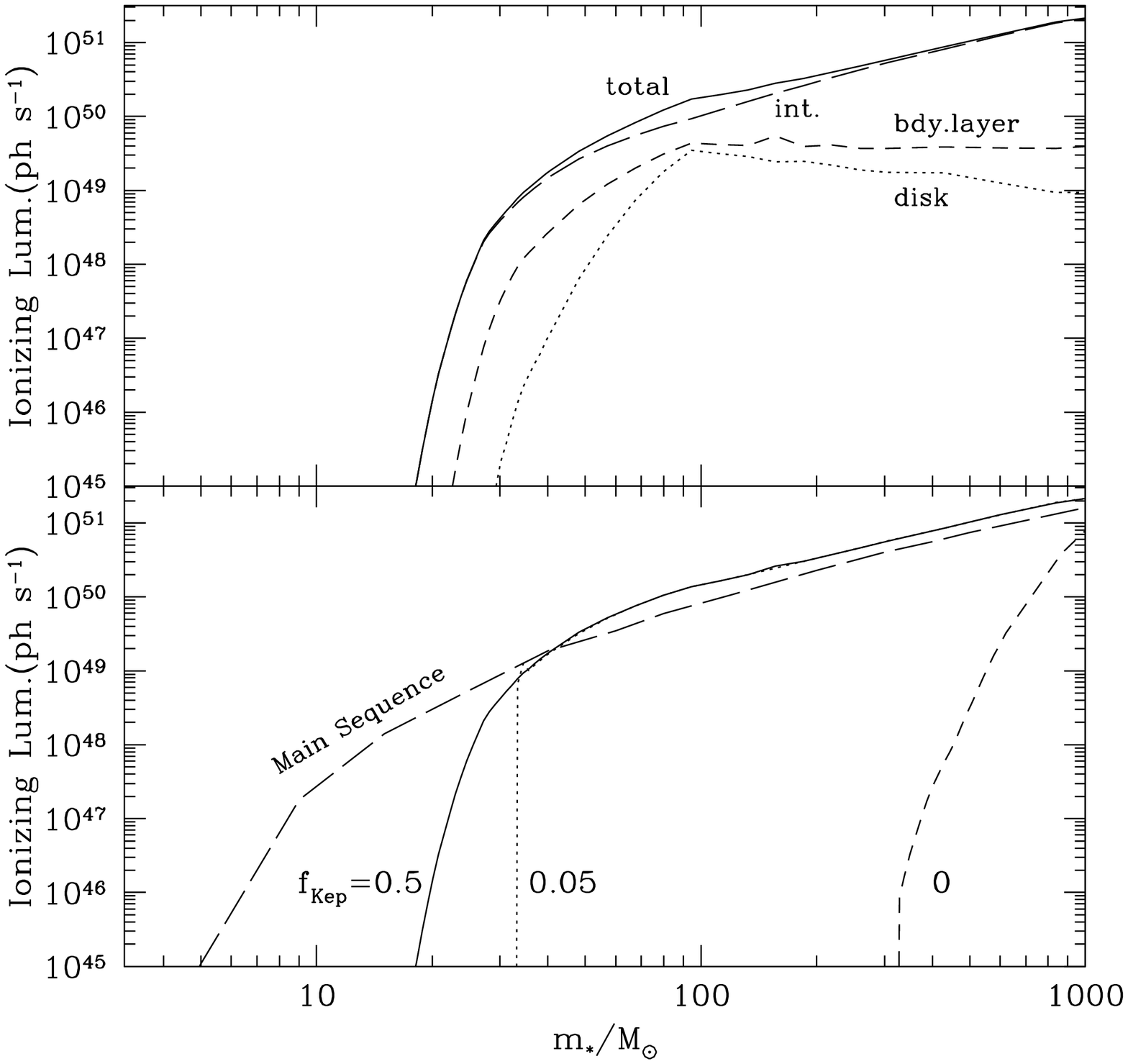,
        angle=0,
        width=6in
}
\end{center}
\caption{
\label{fig:Sstarevolpalla3}
Top: Hydrogen ionizing photon luminosities, $S$, as a function of
protostellar mass for the fiducial isentropic model with rotation
$f_{\rm Kep}=0.5$, $K'=1$, and $\alpha_{\rm ss}=0.01$. The total
ionizing luminosity (solid line) and contributions from the
protostellar surface (long-dashed line), boundary layer (dashed line),
and accretion disk from $r<10r_*$ (dotted line) are shown.  Bottom:
Total hydrogen ionizing photon luminosities for the fiducial spherical
isentropic model (dashed), and with rotation $f_{\rm Kep}=0.05,0.5$
(dotted, solid).  The zero age main sequence ionizing luminosity from
the models of Schaerer (2002) is shown by the long-dashed line.  This
figure shows the crucial importance of the initial core angular
momentum on the evolution of the ionizing luminosity: rotating systems
form protostars that achieve quite high ionizing luminosities at
relatively early stages in their evolution, compared to non-rotating,
spherically-accreting systems.  }
\end{figure}

The ionizing luminosities for different models are shown in Figure
\ref{fig:Sstarevolpalla3}. The contribution from the protostellar
photosphere is usually dominant. The differences between the low and
high rotation parameter cases are very large, reflecting the different
photospheric temperatures. These differences will prove crucial for
determining mass limits to the star formation process from
protostellar feedback (Paper II). Note that in the fiducial model the
ionizing luminosity shows a very rapid increase from $m_*\simeq 20\sm$
to $m_*\simeq 40\sm$, reflecting the dramatic contraction of the star
and the rapidly increasing internal luminosity. For these accretion
rates ($K'=1$) the ionizing luminosity cannot increase much more
quickly than this model, which applies to any rotating model in which
the stellar photosphere coincides with its accretion surface. Thus
without having completed the more detailed calculations of the effects
of feedback on the accretion flow (Paper II), we expect that feedback
processes will not become important until $m_*$ is at least $\sim
30\sm$, and this is an anticipated lower-limit to the mass of the
first stars.

\section{Conclusions}

Recent numerical studies have indicated that the initial conditions
for primordial star formation are dense, massive gas cores in
approximate hydrostatic and virial equilibrium. These physical
properties are set by the microphysics of $\rm H_2$ cooling and not by
the initial cosmological density perturbations. We have presented a
theoretical model for star formation from these cores, basing our
fiducial case on the core formed at the end of the simulation of
Abel, Bryan, \& Norman (2002). Following the philosophy of Stahler,
Shu, \& Taam's (1980) study of low-mass, contemporary star formation,
our approach has been to identify and isolate the various key stages
that can be treated analytically and semi-analytically, and then
combine them into a coherent picture of the whole formation process.
This will be extended to include feedback processes in Paper II.

We have described the rate of collapse of the cores as a function of
the entropy parameter of the gas, $K\equiv P/\rho^{\gamma}$, and the
amount of mass that has already collapsed.  The mass of gas in
hydrostatic equilibrium inside the point at which $T\sim 300$~K, the
minimum temperature due to H$_2$ cooling and $n\sim 10^4$~cm$^{-3}$,
the critical density of $\rm H_2$ rotational-vibrational line cooling,
yields total core masses of about $10^3\sm$. These cores collapse from
the inside-out to form a protostar, which then grows rapidly in mass,
with an accretion rate $\mds \simeq 0.017 (m_*/\sm)^{-3/7}\smyr$.

We have developed a simplified method for modeling protostellar
evolution and applied the appropriate accretion rate for primordial
protostars. The method allows for the treatment of accretion of gas
with angular momentum. Specifically we assume angular momentum is
conserved inside the sonic point of the flow. Using a realistic degree
of rotation for the initial gas core most of the accretion occurs via
a disk and conditions at the protostar rapidly become optically thin,
in contrast to the spherical case. This means that the radiation field
that the accretion envelope is exposed to is significantly hotter so
that ionization and FUV radiation pressure feedback may become
important at much smaller stellar masses than in the spherical case
(Paper II).

In order to determine the emission spectrum of the protostar, we
evaluated the spectrum of the radiation emitted from the stellar
surface in both the optically thick and thin cases, and we included
the emission from the inner accretion disk, including the boundary
layer.  
In modeling the disk, we allowed for the dissociation/ionization
energy of the accreting gas, which can significantly lower the
temperature of the disk.
These predictions will be used in Paper~II to estimate the
importance of various radiative feedback processes.

One may ask how the initial conditions and thus accretion rates of
primordial star formation differ from those of contemporary star
formation. There are several major differences: (1) the mass scale of
cores is higher, being set by the microphysics of $\rm H_2$ cooling,
rather than CO and dust cooling; (2) the gas temperatures are higher,
which follows from (1) and virial equilibrium, so that at the onset of
dynamical collapse the quasi-equilibrium state is denser, the
free-fall time shorter, and the accretion rate larger; (3) because the
accretion rate is larger, primordial protostars are larger and reach
the main sequence at higher masses; (4) correspondingly, support by
deuterium burning is not particularly important, in contrast to
protostars today; (5) thermal pressure is dominant in the initial gas
cores, which is not true for contemporary massive cores (McKee \& Tan
2002), primarily because metal-rich, contemporary cores are permeated
by dynamically important magnetic fields and are able to radiate away
most of their thermal energy, leaving themselves in a state of
non-thermal pressure support that is supersonically turbulent; and (6)
the temperature structure is such that the gas temperature decreases
with radius, leading to accretion rates that decline with time, again
opposite to the evolution of contemporary massive cores.

The contraction to the main sequence (after the star is approximately
older than its Kelvin-Helmholz time), marks a major transition when
radiative feedback from the protostar becomes much more important.
Thus, even before a detailed analysis of these processes (Paper II),
we anticipate that the masses of the first generation of stars must
have been at least as great as $\sim 30\sm$.

\acknowledgements We thank T. Abel, V. Bromm, R. Cen, B. Draine, 
J. Goodman, A. Loeb, E. Komatsu, D. McLaughlin, K. Omukai, and J. Ostriker
for helpful discussions.  JCT is
supported by a Spitzer-Cotsen fellowship from Princeton University and
NASA grant NAG5-10811. The research of CFM is supported by NSF grant
AST-0098365 and by a NASA grant funding the Center for Star Formation
Studies.

\appendix

\section{Energy Accretion onto a Protostar}

Let \beq \epsilon\equiv \frac 12 v^2 +\epsth+\epsilon_I \eeq be the
total energy per gram of gas, where $\epsth=\frac 32 kT/\mu$ is the
thermal energy per unit mass and $\epsilon_I$ is the internal energy per unit
mass, including chemical binding energy.  The zero point of
$\epsilon_I$ is determined by the boundary conditions.  To relate this
to the work of Nakano et al. (1995, 2000), we write
$\epsilon_I=\Psi_I/\mh$.  If the gas is fully molecular, then since it
takes 4.48 eV to dissociate H$_2$, 13.6 eV to ionize H, and 79 eV to
fully ionize He, the maximum value of $\Psi_I$ is $\Psi_{Im}=16.8$ eV
for a primordial gas with 0.079 He per H.

      Conservation of energy, including radiation
but neglecting viscosity and magnetic fields, 
is given by
(see eqs. 94.15b and 96.17 in Mihalas \& Weibel-Mihalas 1999)
\beq
\pbpt \left(\rho\epsilon +\urad\right)
+\vecnabla\cdot\left[\rho\vecv\left(\epsilon+\frac{P_g}{\rho}\right)+\vecF
\right]=\rho\vecg\cdot\vecv+\rho\epsnucd,
\label{eq:encons'}
\eeq
where $\urad$ is the energy density of radiation, $P_g$ is
the gas pressure, $\vecF$ is the flux of radiation measured
in the frame of the protostar, and $\rho\epsnucd$ is the rate of
nuclear energy generation per unit volume.
This equation is accurate to order
$v/c$.
Now, the rate of work by the gravitational field can be
written
\beq
\rho\vecg\cdot\vecv=-\vecnabla\cdot(\rho\vecv\phi)-\pbpt(\rho\phi)
+\rho\frac{\partial\phi}{\partial t}
\eeq
(Shu 1992), so that equation (\ref{eq:encons'}) becomes
\beq
\pbpt \left[\rho(\epsilon+\phi) +\urad\right]
+\vecnabla\cdot\left[\rho\vecv\left(\epsilon+\frac{P_g}{\rho}+\phi
\right)+\vecF
\right]=\rho\frac{\partial\phi}{\partial t}+\rho\epsnucd.
\label{eq:encons}
\eeq

        The rate at which the protostar accretes energy
can be evaluated by integrating equation (\ref{eq:encons}) over
the volume of the protostar. 
Let $E_g$ be the total gas
energy, etc. Then, since the gravitational energy
satisfies
\beq
\frac{dE_{\rm gr}}{dt}=\frac 12\int\pbpt(\rho\phi) dV
              =\int\rho\pbpt(\phi) dV,
\eeq
(Shu 1992), the total energy $E\equiv E_g+E_{\rm gr}+E_r$ obeys
\beq
\frac{dE}{dt}
        +\int{\bf dA\cdot}\rho{\bf v}
        \left(\epsilon+\frac{P_{g}}{\rho}+\phi\right)+L
        =\enucd,
\label{eq:dedtgen}
\eeq
This equation is valid for an arbitrary mass distribution.
The volume of integration must include all the self-gravitating
mass, so it is valid only for $r\ga r_*$.

     We now assume that the protostar is approximately
spherical, so that for $r\ga r_*$ the potential
is
\beq
\phi=-\frac{Gm_*}{r}=-\frac 12 \vff^2,
\eeq
where $\vff$ is the free-fall velocity.
We relate the velocity of the gas to the free-fall velocity
by
\beq
f_k\equiv \frac{v^2}{\vff^2}.
\eeq
Inside the accretion shock, 
$f_k\simeq 0$ since we assume that the star
is not rapidly rotating; in the boundary layer of 
an accretion disk, $f_k\simeq 1/2$; and
in free-fall collapse, $f_k=1$.
Denote the average over the surface of some quantity
$x$ by $\langle x\rangle$, 
so that if the protostar accretes matter directly via free-fall collapse
at a rate $\dot m_{*,\,\rm direct}$ and via disk accretion at
a rate $\dot m_{*,\,\rm disk}$, then
\beq
\langle x\rangle=\frac{1}{\mds}\left( \dot m_{*,\,\rm direct}x_{\rm
        direct}+\dot m_{*,\,\rm disk}x_{\rm disk}\right).
\eeq
Under the assumption that the ratio of specific heats for the
accreting gas is 5/3, 
equation (\ref{eq:dedtgen}) for the energy
becomes
\beq
\frac{dE}{dt}=\mds\left\langle\frac 53\epsth+\epsilon_I
        -\frac 12 (1-f_k)\vff^2\right\rangle-L+\enucd\equiv \mds 
        \langle w\rangle -L+\enucd,
\label{eq:dedt}
\eeq
where $\mds$ is positive for accretion and $w$ is the enthalpy
plus the potential energy per unit mass. 
Note that in contrast to Nakano et al. (1995, 2000), we do not include
nuclear binding energy in our expression for the energy,
and as a result the term $\enucd$ appears on the right-hand side
of equation (\ref{eq:dedt}).

        Since the energy of the protostar is dominated by
its interior, we are free to choose any surface layer at
which to evaluate $dE/dt$.  Equivalently, we can evaluate
equation (\ref{eq:encons}) for steady flow in a region
in which $\epsnucd=0$.  Either way, we conclude that
the luminosity is given by
\beq
L=L_0+\mds\left\langle\frac 53\epsth+\epsilon_I
-\frac 12 (1-f_k)\vff^2\right\rangle=L_0+\mds \langle w\rangle,
\label{eq:L}
\eeq
in terms of $L_0$, the emergent luminosity.
This equation is similar to equation (96.18) of Mihalas \& Weibel-Mihalas
(1999), except that (1) we are focusing on the surface layers so
that there is no nuclear energy generation and (2) we have
not assumed spherically symmetric flow.

\subsection{Energy Flow in an Accretion Disk}

        It is customary to model accretion disks with an anomalous
viscosity $\eta$, although in many cases the actual transport of
energy and angular momentum is mediated by magnetic fields 
(Balbus \& Hawley 1998).  The energy flux due to
viscous transport is $-\vecv\cdot\spr$, where $\spr$ is the viscous
stress tensor.
The energy conservation equation
(eq. \ref{eq:encons}) becomes
\beq
\pbpt \left[\rho(\epsilon+\phi) +\urad\right]
+\vecnabla\cdot\left[\rho\vecv\left(\epsilon+\frac{P_g}{\rho}+\phi
\right)-\vecv\cdot\spr +\vecF
\right]=\rho\frac{\partial\phi}{\partial t}+\rho\epsnucd
\label{eq:enconsvis}
\eeq
(see Landau \& Lifshitz 1959).  
In a steady, axisymmetric, thin (so that $v_\phi^2=Gm/r$, 
$v_z=0$ and $\vecnabla\cdot
\vecF\simeq \partial F/\partial z$) disk in which 
$v_r^2 \ll v_\phi^2$ (so that $v^2\simeq v_\phi^2$ and
$\sigma'_{r\phi}$ is the dominant
stress) and in which 
there are no nuclear reactions, this equation reduces to
\beq
\frac{1}{r}\frac{\partial}{\partial r}r\left[\rho v_r\left(\frac 53
        \epsth +\epsilon_I - \frac
        12\;\frac{Gm}{r}\right)-v_\phi\sigma'_{r\phi}\right]
        +\frac{\partial F}{\partial z}=0,
\eeq
where again we have assumed that the ratio of specific heats
of the gas is 5/3.  Integrating through the thickness of 
the disk, we have
\beq
2F=\frac{1}{r}\frac{\partial}{\partial r}\left[\frac{\dot m}{2\pi}
        \left(\frac 53\bar\epsth +\bar\epsilon_I - \frac
        12\;\frac{Gm}{r}\right)+r v_\phi W_{r\phi}\right],
\label{eq:twof}
\eeq
where $\dot m=-2\pi r\int_{-\infty}^\infty \rho v_r\, dz$ is the accretion
rate (defined to be positive for accretion) and
$W_{r\phi}\equiv \int_{-\infty}^\infty \sigma'_{r\phi} \,dz$ 
is the integrated stress.
We have defined 
\beq
\bar\epsilon\equiv
\frac{1}{2\Sigma_c}\int_{-\infty}^{\infty}\rho\epsilon \, dz
\eeq
as the vertically averaged energy
per unit mass through the disk, where $\Sigma_c\equiv\int_0^\infty
\rho\, dz$ is half of the surface density of the disk.

        From the $\phi$ component of the equation of motion, one
finds that for a Keplerian accretion disk
\beq
W_{r\phi}=-\frac{\dot m\Omega}{2\pi}\left[1-\left(\frac{r_0}{r}\right)^{1/2}
        \right]
\label{eq:wrphi}
\eeq
(Shakura \& Sunyaev 1973), where $\Omega$ is the angular velocity and
$r_0$ is the radius at which
the stress ($\sigma'_{r\phi}=\eta r\partial\Omega/\partial r$,
where $\eta$ is the viscosity coefficient)
vanishes.  
Tan \& Blackman (2004) have argued that a disk dynamo
will create a large enough magnetic field that the
magnetorotational instability (MRI) can be effective
in transporting angular momentum in the disk.
If, however, the magnetic field in the disk is
too weak for the MRI to occur, two caveats must
be kept in mind.  First, in the absence of the MRI,
it is possible that the dominant mode of angular momentum
transport will be due to gravitational torques in
the disk, and it is not clear that the effect of 
these torques can be described even approximately
by a viscous accretion disk. Second, equation
(\ref{eq:wrphi}) is based on the assumption
that the stress vanishes near the surface of
the protostar, which is appropriate if it
is slowly rotating and the transition from
disk to star occurs in a thin boundary layer.
However, for a 
protostar that forms through disk accretion 
in the absence of a magnetic field, it is not clear that there
is a point at which $\partial\Omega/\partial r=0$; instead,
the rotation may approach that of a rigid body at small
radii, becoming very sub-Keplerian without ever passing
through a maximum.  In that case, the value of $r_0$ is
uncertain, but is presumably significantly less than
the stellar radius.

        Inserting equation (\ref{eq:wrphi}) into 
equation (\ref{eq:twof}), 
we then find that the flux emitted from the disk is
\beq
F=\frac{\dot m}{4\pi r}\frac{\partial}{\partial r}\left(\frac 53\bar\epsth +
        \bar\epsilon_I\right)+\frac{3Gm\dot m}{8\pi r^3}
        \left[1-\left(\frac{r_0}{r}\right)^{1/2}\right].
\eeq
The first term is often neglected in discussions of accretion
disks (e.g., Shakura \& Sunyaev 1973; Frank et al. 1995), 
but it can be important in protostellar disks.  For
thin disks, the thermal term $\epsth$ is negligible, but
the ionization term can be up to an order of magnitude larger
than the thermal term and can be significant.

\section{Accretion Directly onto the Protostar}
\label{S:direct}

        Gas that falls directly onto the surface of the protostar
goes through an accretion shock, which we
identify with the surface of the protostar.
The accretion shock has a complex structure:
the shock front, in which the gas temperature jumps
to a high value; the post-shock relaxation layer, in 
which the gas cools by emitting radiation; and
the radiative precursor, in which the gas upstream
from the shock front absorbs and re-radiates the emission
from the post-shock relaxation layer (McKee \& Hollenbach 1980).
The gas and radiation are thus far from LTE inside the accretion
shock.  
Label the surface just outside
the accretion shock front by ``1", and that just inside 
the postshock relaxation layer by ``2".
Stahler et al. (1980) have shown that the Rosseland
mean opacity of the gas between
$r_1$ and $r_2$ is small, so that the radiation pressure 
is the same at both points ($P_{\rm rad, 1}=P_{\rm rad, 
2}$). 
Since the radiation emitted 
by the postshock relaxation layer is 
approximately the
same in the upstream and downstream directions, it follows
that 
the ratio of the mean intensity $J$ to $P_{\rm rad}$
for the radiation emitted in this layer
is the same on
both sides of this layer; because the layer is
optically thin, the same is true for radiation emitted
outside the layer and 
therefore $J_1\simeq J_2$.

     The luminosity in the outer layers of the protostar
varies as
\beq
L=L_0+\mds w,
\label{eq:Lw}
\eeq
where $L_0$ is the emergent luminosity, as seen far from the protostar, and
\beq
w\equiv \frac{5kT}{2\mu}+\epsilon_I-\frac 12 (1-f_k)\vff^2
\eeq
(eq. \ref{eq:L}).  Here $\mu$ is the mean mass per
particle at $r$ and $f_k\equiv v^2/\vff^2$.
For accretion onto the star, $f_{k1}=1$ since 
the unshocked gas is in free-fall collapse and $f_{k2}\simeq 0$
since the shocked gas is nearly at rest.
There is thus
a jump in luminosity across the shock,
\beq
L_1=L_2+\frac 12 f_{k1}\vffs^2+\frac 52\left(\frac{kT_1}{\mu_1}
              -\frac{kT_2}{\mu_2}\right)+\epsilon_{I1}-\epsilon_{I2}.
\label{eq:L1}
\eeq
For free-fall collapse, $L$ decreases beyond $r_1$ as
$\epsth$ and $\epsilon_I$ decrease. 

\subsection{Optically Thin Accretion}
\label{S:optthin}

         To evaluate the conditions at the accretion shock,
we must distinguish between the cases in which the accretion
flow is optically thick or thin 
to photospheric photons; we assume that
the flow is opaque to the energetic photons
emitted by the shocked gas (cf. Stahler et al. 1980).
Let the hot gas behind the shock front emit a flux of energetic
photons $F_x$ upstream and the same flux downstream.
The value of $F_x$ can be inferred from equation
(\ref{eq:L1}), $4\pi r_*^2\cdot 2F_x=L_1-L_2$.
These fluxes are reprocessed into less energetic photons
in the radiative precursor and the postshock relaxation
layer, respectively.  In each case, half the photons
go upstream and half downstream.  (Actually, Calvet \& Gullbring
1998 have shown that the downstream flux from the
radiative precursor can exceed the upstream flux due to
line opacity, but we ignore this complication here.)
As a result, a net 
flux $F_x$ of reprocessed photons enters the postshock
gas at $r_2$. These photons are absorbed and re-emitted
by the shocked gas so that the net flux is zero.  

   In this section, we are assuming that the accretion
flow is transparent. 
We further assume that the radiation emitted by the
hot gas in the postshock relaxation layer and by the
gas in the radiative precursor is significantly more energetic
than that emitted by the photosphere, and that correspondingly the
opacity for the shock photons is significantly greater
than the photospheric opacity (this approximation is
marginal for the reprocessed photons emitted by the
radiative precursor.)
First consider the 
case in which the flux from the interior is
negligible ($\fint\ll F_x$). 
The gas at $r_2$ must then be able to radiate the flux
$F_x$ into space, so that $\sigma T_2^4=F_x$.
In this case the gas inside $r_2$ is isothermal,
so this is also the effective
temperature $T_{\rm eff, 2}$. Including the effect of
$\fint$ increases the effective
temperature of the gas behind the postshock relaxation layer to
\beq
\sigma T_{\rm eff,2}^4=\fint + F_x.
\label{eq:teff2}
\eeq
In the Eddington approximation the contribution
of the interior flux to $\sigma T^4$ is reduced
by a factor 2 at the surface, since the radiation
then occupies only half the available solid angle.
As a result, we have
\beq
\sigma T_2^4=\frac 12 \fint + F_x.
\label{eq:t2}
\eeq
Stahler et al. (1980) evaluated the temperature inside
the postshock relaxation layer, at a point at which the
X-rays had been emitted but not yet reprocessed, and they
obtained a factor $\frac 32$ in front of $F_x$ in this
equation (corrected by Stahler 1988).  The calculations
of Calvet \& Gullbring (1998---see below) show that the temperature
structure of the shocked gas is actually more complicated
than these simple analytic models imply, however.
The effective temperature at $r_1$ includes the
contribution of the reprocessed flux $F_x$ emitted
upstream, so that
\beq
\sigma T_{\rm eff, 1}^4=\fint +2F_x.
\label{eq:teff1}
\eeq

      Calvet \& Gullbring (1998) carried out both
analytic and numerical models of the accretion shock.  
In their analytic calculation, they
divided the postshock relaxation layer into two
parts, the region in which the energetic photons
are emitted and that in which they are absorbed,
and they included the latter region in their calculation.
With a flux $F_i$ incident on the absorbing gas,
they found
\beq
\sigma T^4=\frac 34\fint\left(\tau+\frac 23\right)
       +\frac 14 F_i\left[2+\frac{3}{q}+\left(q-\frac 3q\right)
       e^{-q\tau}\right],
\eeq
where $q$ is the ratio of the opacity of the energetic photons
to the photospheric photons. The first term is the standard
Eddington approximation for the temperature structure of a plane
parallel atmosphere carrying a flux $\fint$.
If $q\gg 1$, as we have assumed,
then the contribution of the shock radiation to the effective
temperature is
$$
\frac 14 F_i\left(2+q e^{-q\tau}\right).
$$
At $\tau=0$ (which is inside the shock in our 
terminology), $T^4$ can be much larger than $F_i$.
Only the reprocessed photons penetrate into the
region $r\leq r_2$.  Since an approximately equal flux
of reprocessed photons enters this region from the
radiative precursor, the solution in this region
is equivalent to having $F_i=2F_x$.
Since the energetic photons have been absorbed at $r_2$
($q\tau\gg 1$) but not photospheric photons ($\tau\ll 1$), it follows
that $\sigma T_2^4=\frac 12 (\fint+ F_i)=\frac 12 \fint +F_x$,
in agreement
with our result in equation (\ref{eq:t2}).

        Calvet \& Gullbring's (1998) solution shows that
the shock is thin only if $q\gg 1$, as we have assumed. In this case,
the gas emitting the reprocessed photons is hotter
than the photosphere by a factor $\sim q^{1/4}$. As a
result, the radiation
outside the photosphere does not have a blackbody
spectrum; correspondingly, the analytic model
for the temperature distribution is approximate. 
However, their numerical calculations show that equation
(\ref{eq:teff2}) for the effective temperature behind the shock
is accurate to within a few percent.  The gas between
the photosphere and the shock is generally warmer than
that at the photosphere due to the heating by reprocessed
photons.

        How does the heating and ionization of the gas ahead
of the shock affect the emitted luminosity?  
The luminosity that emerges from the protostar can
be inferred from equation (\ref{eq:L}) 
\beq
L_0=L_2-\mds\left(\frac{5kT_2}{2\mu_2}+\epsilon_{I2}
        -\frac 12 \vff^2\right),
\eeq
where we have assumed that $L_0$ is measured at a large
enough distance from the star that $T_0$ and $\epsilon_0$ are
both negligible.
To approximately allow for the non-blackbody nature
of the spectrum of the radiation near the shock, we
evaluate the ionization at the effective temperature
behind the shock, $\epsilon_{I2}=\epsilon_I(T_{\rm eff, 2})$.

\subsection{Optically Thick Accretion}
\label{S:optthk}

        When the accretion flow is opaque, we determine
the temperature just behind the accretion shock, $T_2$, from
the radiation momentum equation (Mihalas \& Weibel-Mihalas 1999, eqs. 
97.2 and 97.3),
\beq
{\bf \nabla\cdot P}_{\rm rad}'=-\frac{\kappa{\bf F}'}{c},
\eeq
where ${\bf P}_{\rm rad}'$ is the radiation pressure tensor and
${\bf F}'$ the flux, both measured in the comoving frame,
and where $\kappa$ is the Rosseland mean opacity.
The comoving flux is related to that in the protostar frame
by $F'=F-v(u_{\rm rad}'+P_{\rm rad}')$ 
(Mihalas \& Weibel-Mihalas 1999, eq. 91.17).
Hence, for isotropic radiation that is in LTE ($P_{\rm rad}'
=a_RT^4/3$) we have
\beq
\frac{\partial T}{\partial r}=-\kappa\left(\frac{3 F}{ca_RT^3}
           -\frac{vT}{c}\right).
\label{eq:dtdr}
\eeq
Note that $v<0$ for accretion flow, so the advection term
$\kappa vT/c$ increases the magnitude of the temperature gradient.
To simplify the integration, we approximate the luminosity
as a constant, $L\simeq \frac 12(L_1+L_p)$, where
$L_p$ is the photospheric luminosity. The flux is then
\beq
F\simeq \frac{L_1+L_p}{8\pi r^2}.
\eeq

        To solve for $T_2$, we first guess a value of $L_p$,
which determines the photospheric temperature $T_p$ in terms
of the unknown radius of the photosphere, $r_p$ through
$T_p=(L_p/4\pi \sigma r_p^2)^{1/4}$.
We solve for
the temperature structure and thus the total optical depth of the
radiative precursor from the photosphere to the accretion shock, using
the opacities of Rogers \& Iglesias (1992) and Iglesias \& Rogers
(1996) for $T>7000\:{\rm K}$ and those of Lenzuni, Chernoff, \&
Salpeter (1991) for $T<7000\:{\rm K}$. This is carried out at an angle of
$\pi/3$ from the pole, so that the optical depth is a reasonable
approximation for the mean value from the stellar surface.
According to equation
(\ref{eq:Lw}), $L_p$ and $L_2$ are related by
\begin{eqnarray}
L_p=&&L_2+\frac 12\mds\vffs^2\nonumber\\
%       \left[1-(1-\langle f_{kp}\rangle)\frac{r_*}{r_p}\right]\nonumber\\
         && {} -\frac 52\mds\left(
          \frac{kT_2}{\mu_2}-\frac{kT_p}{\mu_p}\right)
          -\mds(\epsilon_{I2}-\epsilon_{Ip}).
\end{eqnarray}
This value of $L_p$ is used to determine $T_p$ and the cycle is
repeated until convergence is achieved.

%--------------------------------------------------------------------------
\end{document}